\documentclass{aa}  
\usepackage{graphicx}
\usepackage{hyperref}
\usepackage{amsmath,amssymb} 
\usepackage{txfonts}
\usepackage[T1]{fontenc}
\usepackage{afterpage}
\usepackage{amssymb}
\usepackage{natbib} 
\usepackage{wrapfig}
\usepackage{graphicx}
\usepackage{soul}
\usepackage{color, colortbl}
\bibpunct{(}{)}{;}{a}{}{,}
\newcommand{\Ms}{\ensuremath{M_\odot}}

\newcommand{\Dh}{D$_{\rm h}$}
\newcommand{\Dv}{D$_{\rm v}$ }
\newcommand{\Deff}{D$_{\rm eff}$ }

 \def\lesssim{\mathrel{\hbox{\rlap{\hbox{\lower4pt\hbox{$\sim$}}}\hbox{$<$}}}}
 \def\gtrsim{\mathrel{\hbox{\rlap{\hbox{\lower4pt\hbox{$\sim$}}}\hbox{$>$}}}}

\begin{document}

   \title{Anisotropic turbulent transport\\in stably stratified rotating stellar radiation zones}

   \subtitle{}

   \author{S. Mathis\inst{1,2}
          \and
          V. Prat\inst{1,2}
          \and
          L. Amard\inst{3,4,5}
          \and
          C. Charbonnel\inst{4,6}
          \and
          A. Palacios\inst{3}
          \and N. Lagarde \inst{7}
          \and P. Eggenberger \inst{4}
			}

   \institute{IRFU, CEA, Universit\'e Paris-Saclay, F-91191 Gif-sur-Yvette, France\\
              \email{stephane.mathis@cea.fr}
         	\and
                Université Paris Diderot, AIM, Sorbonne Paris Cit\'e, CEA, CNRS, F-91191 Gif-sur-Yvette, France
            \and
 			    LUPM, Universit\'e de Montpellier, CNRS, Place E. Bataillon - cc 072, F-34095 Montpellier Cedex 05, France
            \and
                Department of Astronomy, University of Geneva, Chemin des Maillettes 51, CH-1290 Versoix, Switzerland
            \and
                University of Exeter, Department of Physics \& Astronomy, Stoker Road, Devon, Exeter, EX4 4QL, UK
            \and 
               	IRAP, UMR 5277, CNRS and Universit\'e de Toulouse, 14 Av. E. Belin, F-31400 Toulouse, France            
            \and Institut UTINAM, CNRS UMR 6213, Univ. Bourgogne Franche-Comt\'e, OSU THETA Franche-Comt\'e-Bourgogne, Observatoire de Besançon, BP 1615, 25010, Besançon Cedex, France 
             }

   \date{\today}

 
  \abstract
   {Rotation is one of the key physical mechanisms that deeply impact the evolution of stars.
Helio- and asteroseismology reveal a strong extraction of angular momentum from stellar radiation zones 
over the whole Hertzsprung-Russell diagram.}
   {Turbulent transport in differentially rotating stably stratified stellar radiation zones should be carefully modeled and its strength evaluated. Stratification and rotation imply that this turbulent transport is anisotropic. Only phenomenological prescriptions have been proposed for the transport in the horizontal direction, which however constitutes a cornerstone in current theoretical formalisms for stellar hydrodynamics in evolution codes. We aim at improving its modeling.}
   {We derive a new theoretical prescription for the anisotropy of the turbulent transport in radiation zones using a spectral formalism for turbulence that takes simultaneously stable stratification, rotation, and a radial shear into account. Then, the horizontal turbulent transport resulting from 3D turbulent motions sustained by the instability of the radial differential rotation is derived. 
We implement this framework in the stellar evolution code STAREVOL 
 and quantify its impact on the rotational and structural evolution of solar metallicity low-mass stars from the pre-main-sequence to the red giant branch. 
   }
   {The anisotropy of the turbulent transport scales as $N^4\tau^2/\left(2\Omega^2\right)$, $N$ and $\Omega$ being the buoyancy and rotation frequencies respectively and $\tau$ a time characterizing the source of turbulence
   . This leads to a {horizontal turbulent transport of similar strength in average that those obtained with previously proposed prescriptions even if it can be locally larger below the convective envelope. Hence the models computed with the new formalism still build up too steep internal rotation gradients compared to helioseismic and asteroseismic constraints. As a consequence, a complementary transport mechanism like internal gravity waves or magnetic fields is still needed to explain the observed strong transport of angular momentum along stellar evolution.}}
   {The new prescription 
 links for the first time the anisotropy of the turbulent transport in radiation zones to their stratification and rotation. This constitutes {an important theoretical} progress and demonstrates how turbulent closure models should be improved to get firm conclusions on the potential importance of other processes that transport angular momentum and chemicals inside stars along their evolution.} 


   \maketitle
%

\section{Introduction}

Rotation is one of the key physical mechanisms that deeply modify the dynamics and evolution of stars \citep[e.g.][]{Maeder2009}. The transport of angular momentum and chemicals it induces in their stably stratified radiation zones drives their secular rotational and chemical evolution. Rotation modifies the evolutionary path of stars in the Hertzsprung-Russell diagram (hereafter HRD), 
their life-time, their nucleosynthesis, chemical stratification and yields, and their magnetism.

In this context, helio- and asteroseismology provide key information through the insight they give on the internal rotation profiles of the Sun and stars.
On one hand, helioseismic data show that the radiative core of the Sun is rotating as an almost solid body down to $r=0.2R_{\odot}$ {with a potential central acceleration} \citep{Brownetal89,Thompsonetal2003,Garciaetal2007,Fossatetal2017}. On the other hand, asteroseismic data reveal a strong extraction of angular momentum 
over the whole HRD. 
First, \cite{Becketal2012}, \cite{Mosseretal2012b}, \cite{Deheuvelsetal2012}, \cite{Deheuvelsetal2014}, \cite{Deheuvelsetal2015}, \cite{Spadaetal2016}, and \cite{Gehanetal2018} found a weak core-to-surface rotation contrast in low-mass subgiant and red giant stars.
Next, \cite{Benomaretal2015} observed 26 solar-type stars with a small differential rotation between the base of their convective envelope and the upper part of their radiative core.
In addition, weak differential rotation rates are found in the radiative envelope of intermediate-mass and massive stars \citep{Kurtzetal2014,Saioetal2015,Trianaetal2015,Murphyetal2016,Aertsetal2017}. Finally, a strong extraction of angular momentum is required to explain the rotation rates of white dwarfs \citep[e.g.][]{Setal2008,Hermesetal2017} and neutron stars \citep[e.g.][]{Hegeretal2005,HM2010}.

In his seminal paper, \cite{Zahn1992} proposed for the first time a consistent and complete formalism to describe the secular transport of angular momentum and chemicals under the combined action of rotation-driven vertical and horizontal turbulence and meridional flows \citep[see also][]{MZ1998,MZ2004}. This theoretical treatment of rotation relies on a key physical assumption: an anisotropic turbulent transport in stellar radiation zones stronger in the horizontal (latitudinal) direction than in the vertical one because of the restoring buoyancy force along the radial entropy (and chemical) stratification. This strong horizontal transport erases horizontal gradients of physical quantities, among which angular velocity, thus enforcing a "shellular" rotation depending only on the radial coordinate. This framework allowed for a successful implementation in 1D stellar evolution codes \citep{Talonetal1997,MeynetMaeder2000,Palaciosetal2003,Decressinetal2009,Ekstrometal12,Marquesetal2013,ChieffiLimongi2013} with numerous applications over a broad range of stellar types and evolutionary stages. 

However, disagreements between the predictions of this formalism and seismic data on internal stellar rotation at various stages of the evolution \citep{TZ1998,TCetal2010,Eggenetal2012,Marquesetal2013,Ceillieretal2013} led the community to examine the role of other transport mechanisms for angular momentum, such as internal gravity waves \citep[e.g.][]{TKZ2002,TC2005,CT2005,Charbonneletal13,Rogers2015,Pinconetal2017} and magnetic fields \citep[e.g.][]{GM1998,Spruit1999,MZ2005,Eggenbergeretal05,DP07,SBZ2011,AcevedoGaraudWood2013,Barnabeetal2017}. At the same time, the physical description of shear-induced vertical and horizontal turbulence in stably stratified stellar layers was not questioned. 

The aim of the present paper is to bring a new light on the hydrodynamical processes induced by rotation and on their role in the whole picture, based on the most recent numerical and theoretical developments. 
In this context, recent high-resolution numerical simulations in local Cartesian configurations that were performed to estimate the turbulent transport induced by the instability of a vertical shear \citep{Pratetal2013,Pratetal2014,Pratetal2016,Garaud2016,Garaud17} can be used for guidance. Their comparison with former phenomenological prescriptions for related turbulent transport coefficients in the radial direction \citep{Zahn1992,TZ1997} indeed invites stellar astrophysicists to reconsider this status. 

This is particularly true for turbulence in the horizontal direction. Indeed, while strong stratification is invoked as the source of the strong anisotropic transport \citep{Zahn1992}, none of the currently used prescriptions for horizontal turbulent transport coefficients in stellar evolution codes depends explicitly on the two restoring forces in stellar radiation zones: stratification and rotation. Two mechanisms can be identified to sustain the horizontal turbulent transport: (i) the instability of the shear of the latitudinal differential rotation, but also (ii) the 3D turbulent motions induced by the instability of the radial differential rotation that transport momentum and chemicals both in the vertical and horizontal directions. Symmetrically, \cite{Zahn1992} already identified the two sources for the vertical turbulent transport, i.e., (a) the instability of the shear of the radial differential rotation and (b) the transport induced by the 3D turbulent motions induced by the instability of the latitudinal differential rotation. He introduced the corresponding vertical turbulent transport coefficients, $\nu_{\rm v,v}$ and $\nu_{\rm v,h}$, modelled as eddy-viscosities, and the corresponding eddy-diffusivities, $D_{\rm v,v}$ and $D_{\rm v,h}$. Note that in stellar physics modeling though, the simplifying assumption 
$\nu_{\rm v,v}\equiv D_{\rm v,v}$ and $\nu_{\rm v,h}\equiv D_{\rm v,h}$ has been 
made until now. Symmetrically, 
four coefficients $\nu_{\rm h,h}$, $D_{\rm h,h}$, $\nu_{\rm h,v}$ and $D_{\rm h,v}$ can also be defined for the horizontal transport (Tab. \ref{tab:coefficient} is recapitulating the different turbulent diffusivities (viscosities) and their source). 
On one hand, the first developments focused on the instability of the latitudinal differential rotation (i.e. on $\nu_{\rm h,h}$ and $D_{\rm h,h}$). \cite{Zahn1992} initially proposed a prescription based only on phenomenological arguments. Next, \cite{Maeder2003} derived a prescription based on the evaluation of the dissipation of the energy contained in a horizontal shear. Finally, \cite{MPZ2004}, derived a prescription based on results observed for turbulent transport in a non-stratified Taylor-Couette experiment \citep{RZ1999}. On the other hand, theoretical works \citep[e.g.][]{BC2001,KB2012} and numerical simulations \citep{WB2006,KB2012} in fundamental and astrophysical fluid dynamics have been devoted to characterize key properties of anisotropic turbulent flows in rotating stably stratified media, such as velocities and length scales in the vertical and horizontal directions. This is a good motivation to also consider the horizontal transport induced by 3D turbulent motions sustained by the instability of the radial differential rotation, which has been ignored until now.

In this work, we derive a new prescription for the corresponding turbulent transport coefficients (i.e. $\nu_{\rm h,v}$ and $D_{\rm h,v}$). This allows us to propose for the first time an expression that depends explicitly on stratification, rotation, and their ratio. In Sect.~\ref{sec:fluid}, we generalize the spectral model introduced by \cite{KB2012} to study the anisotropy of turbulent flows in differentially rotating stably stratified layers and we derive scaling laws allowing us to establish our new prescriptions. In Sect.~\ref{sec:appli}, we implement them in the stellar evolution code STAREVOL. We study their impact on the rotational and structural evolution of a 1.0$M_\odot$ star during its pre-main sequence and main-sequence and on the subgiant and giant phases of a 1.25$M_\odot$  star. Finally, we discuss results and perspectives of this work in the conclusion (Sect.~\ref{sec:conclusion}). 

\begin{table*}[ht!]
\begin{center}
\caption{The turbulent diffusivities (viscosities) and their source.}
\begin{tabular}{ c c }
\hline
\hline
$D_{\rm v,v}$ ($\nu_{\rm v,v}$) & Vertical turbulent transport induced by the vertical shear instability \\
$D_{\rm h,v}$ ($\nu_{\rm h,v}$) & Horizontal turbulent transport induced by the 3D motions of the vertical shear instability \\
$D_{\rm h,h}$ ($\nu_{\rm h,h}$) & Horizontal turbulent transport induced by the horizontal shear instability \\
$D_{\rm v,h}$ ($\nu_{\rm v,h}$) & Vertical turbulent transport induced by the 3D motions of the horizontal shear instability\\
\hline
\end{tabular}
\label{tab:coefficient}
\end{center}
\end{table*}

\section{Turbulent transport in stably stratified differentially rotating layers}
\label{sec:fluid}

\subsection{Theoretical background}

\subsubsection{Stably stratified fluids}

Stably stratified stellar radiation zones are supposed to be the place 
of a mild mixing induced by rotation that deeply impacts the evolution of stars \citep[e.g.][]{Zahn1992,Maeder2009}. Among the physical processes at the origin of such secular transport of chemicals and angular momentum, shear-induced turbulence plays a key role \citep{KS1982,Zahn1983,Zahn1992,TZ1997,MPZ2004}.

\begin{figure}
\centering
\label{Fig:Anisotropy}
\includegraphics[width=0.35\textwidth]{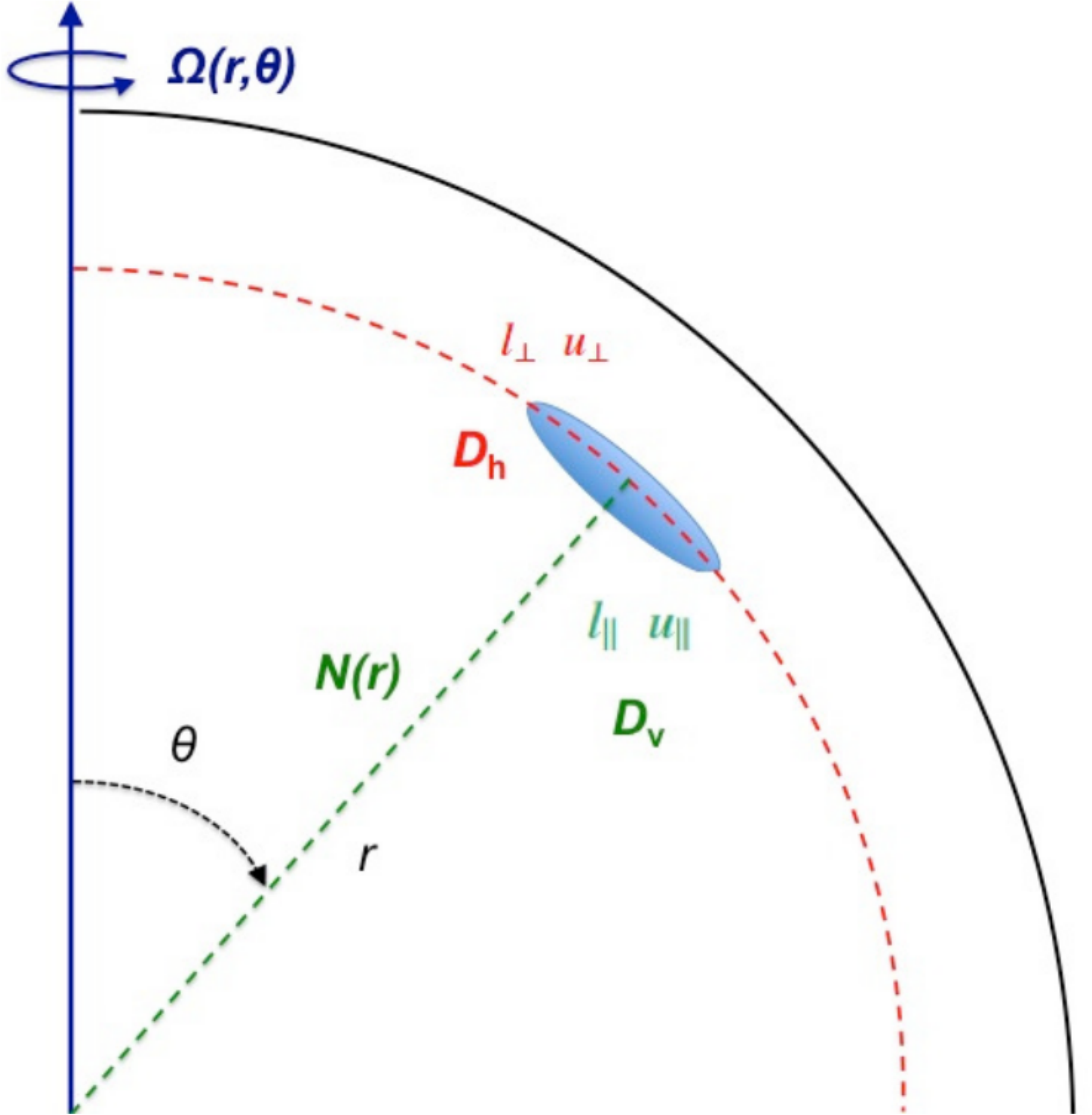}
\caption{Anisotropic turbulent transport in a stably stratified stellar radiation zone with a buoyancy frequency $N$ rotating with an angular velocity $\Omega$. We introduce the characteristic velocities ($u_{\parallel},u_{\perp}$) and length scales ($l_{\parallel},l_{\perp}$) of a turbulent "pancake" in the direction of the entropy stratification and in the horizontal one respectively and usual spherical coordinates ($r,\theta$)}
\label{Scheme}
\end{figure}

In this context, the impact of stable stratification on turbulent motions that develop in radiative regions must be understood in detail and carefully evaluated. Indeed, the motions become highly anisotropic because of the potentially strong buoyancy force along the direction of the entropy and mean molecular weight stratifications \citep[e.g.][and references therein]{RL2000,BC2001,Brethouweretal2007,Davidson2013,Marinoetal2014}. It is quantified by the Brunt-V\"ais\"al\"a frequency $N$, defined by
\begin{equation}
N^2=
\frac{g \delta}{H_{P}}\left(\nabla_{\rm ad}-\nabla\right),
\label{BVf}
\end{equation}
where we consider only the thermal part with 
$g$ the gravity, $H_{P}=\vert{\rm d}r/{\rm d}\ln P\vert$ the pressure scale height, and $P$ the pressure. We introduce the logarithmic gradients $\nabla={\rm d}\ln T/{\rm d}\ln P$ with $T$ the temperature,
and $\nabla_{\rm ad}=\left({\partial}\ln T/{\partial}\ln P\right)_{\rm ad}$, and the thermodynamic coefficient $\delta=-\left(\partial\ln\rho/\partial\ln T\right)_{P,\mu}$ 
where $\rho$ is the density and $\mu$ the mean molecular weight. The component of the buoyancy frequency related to $\mu$-gradients is here ignored as a first step. Buoyancy force reduces the amplitude of the vertical displacement of turbulent eddies along the direction parallel to the stratification (hereafter denoted $\parallel$) while no restoring force is applied along the perpendicular (generally horizontal) direction (hereafter denoted $\perp$), if ignoring the action of the Coriolis acceleration (see Sect.~\ref{subsec:rotation}). This anisotropy induces turbulent velocities $u_{\parallel}$ and $u_{\perp}$ in the directions parallel and perpendicular to the stratification respectively. Moreover, flows become organized as horizontal "pancake-like" structures with characteristic vertical and horizontal length scales, $l_{\parallel}$ and $l_{\perp}$ (see Fig.~\ref{Scheme}).
When occurring in stellar radiation zones, such an anisotropic turbulence can have important consequences \citep{Zahn1992}. Indeed, because of the action of the buoyancy force in the vertical direction and of the lack of restoring force in the horizontal one, it would lead to a stronger transport of angular momentum and chemicals horizontally than vertically. This may lead to only weak horizontal gradients of angular velocity ($\Omega$), entropy (and temperature) and mean molecular weight \citep{ChaboyerZahn1992}. As proposed by \cite{Zahn1992}, rotation then can become "shellular" with related simplifications of 2D transport equations \citep{Zahn1992,MZ1998,MZ2004} that have been successfully implemented in several stellar evolution codes: the Geneva code \citep[e.g.][]{Talonetal1997,MeynetMaeder2000,Eggenbergeretal08,Ekstrometal12}, the code STAREVOL \citep[e.g.][]{Palaciosetal2003,Decressinetal2009,CCNL10,Amardetal2016}, the code CESTAM \citep[e.g.][]{Marquesetal2013}, and the code FRANEC \citep[e.g.][]{ChieffiLimongi2013}.

The control parameter of such dynamics is the ratio between the vertical and horizontal turbulent transport coefficients
\begin{equation}
D_{\rm v}\equiv D_{\parallel}\propto u_{\parallel}\,l_{\parallel}=u_{\parallel}^{2}\tau=\frac{l_{\parallel}^{2}}{\tau}
\end{equation}
and
\begin{equation}
D_{\rm h}\equiv D_{\perp}\propto u_{\perp}\,l_{\perp}=u_{\perp}^{2}\tau=\frac{l_{\perp}^{2}}{\tau},
\end{equation}
where $\tau$ is a dynamical time scale characterizing the turbulence and its source, when assuming a diffusive description for each of them \citep{ChaboyerZahn1992}. Such an assumption is justified when the effect of the instabilities in their non-linear regime is to cancel their source \citep{RZ1999,MPZ2004}.
Then, their ratio can be written
\begin{equation}
\label{eq:ratio}
\frac{D_{\rm v}}{D_{\rm h}}=\frac{u_\parallel^2}{u_\perp^2}=\frac{l_\parallel^2}{l_\perp^2}=\varepsilon^2,
\end{equation}
where we introduce $\varepsilon= l_\parallel/l_\perp$. From now on to \S 2.2.2., we lighten notations by using $\left\{D_{\rm v},D_{\rm h}\right\}$ that stands for $\left\{D_{\rm v,v},D_{\rm h,v}\right\}$ or $\left\{D_{\rm v,h},D_{\rm h,h}\right\}$ depending on the studied source of turbulence, namely the vertical and horizontal shear, respectively.\\
To go further, scaling laws can be obtained using different approaches. Following \cite{BC2001} \citep[see also][]{Davidson2013}, we can introduce the horizontal Froude number of the flow
\begin{equation}
Fr_{\perp}=\frac{u_{\perp}}{l_{\perp}N}.
\end{equation}
It compares the relative strength of the inertia associated with the horizontal advection of the horizontal turbulent motions and buoyancy \footnote{The Froude number is the equivalent for stably stratified flows of the Rossby number for rotating flows.}. It can also be interpreted as the ratio between the time associated to internal gravity waves ($\tau_{\rm IGWs}=1/N$) and the turbulent characteristic time $\tau$. In their approach, \cite{BC2001} consider $u_{\perp}$ and $l_{\perp}$ as fixed quantities. The vertical characteristic velocity ($u_{\parallel}$) and length scale ($l_{\parallel}$) result from the dynamics of the stably stratified flow. Using dimensionless quantities in the Navier-Stockes, continuity, and heat transport equations (in the inviscid and adiabatic limits) describing the dynamics of non-rotating stably stratified flows (we refer the reader to the Eqs. (7-10) of their article), they showed that a second vertical Froude number 
\begin{equation}
Fr_{\parallel}=\frac{Fr_{\perp}}{\varepsilon}=\frac{u_{\perp}}{l_{\parallel}N}
\end{equation}
is also a characteristic number of the system. It compares the inertia of the vertical advection of the horizontal turbulent motions and buoyancy. Making an asymptotic expansion of these equations in the regime of strong stratification for which $Fr_{\perp}\!<\!\!<\!1$, they identified a self-similarity that leads to
\begin{equation}
\varepsilon=\frac{l_{\parallel}}{l_{\perp}}\approx\frac{u_{\parallel}}{u_{\perp}}\approx Fr_{\perp}\,\hbox{and thus}\, Fr_{\parallel}\approx 1. 
\end{equation}
These scaling laws have been verified in high-resolution non-linear Cartesian numerical simulations computed by \citep{Brethouweretal2007} for high Reynolds numbers.

This implies that
\begin{equation}
\frac{D_{\rm v}}{D_{\rm h}}\approx Fr_\perp^2.
\label{Anis1}
\end{equation}
Therefore, as expected, the anisotropy becomes larger when stable stratification is stronger because of the corresponding vertical buoyancy restoring force. Thus, $D_{\rm v}/D_{\rm h}\rightarrow 0$ (or $D_{\rm h}/D_{\rm v}\rightarrow \infty$) when $N\rightarrow\infty$.
However, stars are potentially rapidly rotating, e.g. young solar-type stars and intermediate-mass and massive stars \citep[e.g.][]{MeynetMaeder2000,GB2013,GB2015}. Moreover, their angular velocities vary over several orders of magnitude along their evolution. It is thus mandatory to understand and quantify the effect of the Coriolis acceleration, which will also constrain turbulent motions, particularly in the horizontal direction.

\subsubsection{The effects of rotation}

The effects of the stable stratification in the vertical direction on turbulent flows being now introduced, the balance in the horizontal direction and the effects of rotation through the Coriolis acceleration should be examined. Indeed, if turbulent motions induced by the vertical shear instability are considered, they are intrinsically three-dimensional and they induce turbulent transport and mixing both in the vertical direction, where they are submitted to the buoyancy force, and in the horizontal direction, where the Coriolis acceleration is the restoring force. 

In his seminal paper, \cite{Zahn1992} already pointed out the key role of rotation to control the anisotropy of turbulent transport in stably stratified stellar radiation zones. He discussed the transition between 3D isotropic to 2D rotationally-controled turbulent flows \citep[see e.g.][for experimental studies]{Hopfingeretal1982} and its consequence for the vertical turbulent transport. In this context, he also showed how the instability of an horizontal shear contributes to this transport in the radial direction and he derived the prescriptions for the eddy-viscosity $\nu_{\rm v,h}$ \citep[we refer the reader to Eqs. 2.19, 2.20, 2.22 and 2.23 in][]{Zahn1992}. This already points out how the tridimensional turbulence induced by the instability of a shear varying along a given direction (radial or latitudinal) sustains turbulent transport both in the vertical and horizontal directions. Therefore, one must take into account the contributions of both vertical and horizontal shear instabilities when studying the horizontal (and vertical) turbulent transport in stellar radiation zones.

In addition, several works in fundamental fluid mechanics examined the strength of the anisotropy of turbulence in stably stratified rotating flows as a function of their buoyancy frequency ($N$) and angular velocity ($\Omega$). Depending on the rotation rate, \cite{BC2001} proposed that the anisotropy is driven either mainly by stratification (in the regime of "slow" rotation) or by the combined action of stratification and rotation (in the regime of "rapid" rotation). The physical parameter that controls the transition from one regime to the other is the dimensionless Rossby number 
\begin{equation}
Ro^{\rm h}=\frac{u_{\perp}}{2\Omega l_{\perp}}, 
\end{equation}
which quantifies the relative strength of advection in the horizontal direction to the Coriolis acceleration. In the slowly rotating regime ($Ro^{\rm h}\rightarrow \infty$), their formalism recovers results discussed in the previous section when ignoring rotation. In the rapidly rotating regime ($Ro^{\rm h}\!<\!\!<\! 1$), which is of interest for stellar radiation zones\footnote{See Appendix~\ref{appendix:shellular}}, they found that  
\begin{equation}
l_{\perp}=\frac{N}{2\Omega}l_{\parallel}.
\end{equation}
We here recognize the internal Rossby deformation radius in the case where the full Coriolis acceleration is taken into account \citep[][see also \citealt{Pedlosky1982}]{DS2005}. It characterizes the restoring action of the Coriolis acceleration in the horizontal direction, along which rotation limits the size of turbulent stratified pancakes with $l_{\perp}$ decreasing with increasing $\Omega$. Therefore, rotation competes with stratification, which sustains the anisotropy of turbulent flows with $l_{\perp}$ that increases with $N$. These theoretical scaling laws have been confirmed by \cite{WB2006} who computed high-resolution non-linear numerical simulations in Cartesian geometry where they explored the variations of $l_{\parallel}$ as a function of $Ro^{\rm h}$. This proposed scaling in conjonction with Eq.~\eqref{eq:ratio} leads to
\begin{equation}
\frac{D_{\rm v}}{D_{\rm h}}\approx\left(\frac{2\Omega}{N}\right)^{2}.
\end{equation}
However, these studies do not take into account a large-scale shear, the key physical ingredient of stellar radiation zones we are studying here.

In this framework, the results obtained by \cite{KB2012} are of great interest. Indeed, although they focused in their work only on the case of uniform rotation as in \cite{BC2001}, they derived a general analytical spectral formalism that allows one to derive the response of a stratified rotating fluid to a pre-existing source of turbulence and the anisotropy of the induced turbulent transport. In the cases they studied, they compared the predictions of this turbulent model to direct high-resolution non-linear numerical simulations and showed that the results agree very well. This strongly motivates a generalization of their work by taking into account simultaneously a large-scale shear, stable stratification and rotation to study the problem of anisotropic turbulent transport induced by shear instabilities in stellar radiation zones. This is what we propose below.


\subsection{Anisotropic turbulent transport in stably stratified rotating radially sheared flows}

\subsubsection{Spectral formalism}

In this section, we generalize the formalism of \citet{KB2012} in the case of a sheared, rotating, stably stratified flow. The idea is to determine the response of the fluid to a pre-existing source of turbulence characterized by a velocity field $\vec u^0$. We write fluid equations using the Boussinesq approximation where we assume that space-scales characterizing turbulent motions are smaller than those of the variations of background quantities. 

Moreover, following \cite{KB2012}, a relaxation approximation is used to get rid of non-linear terms.
It consists in assuming that the fluid tends to reach a steady state with a typical time scale $\tau$.
Mathematically, it reads
\begin{equation}
	\frac{{\rm d}X}{{\rm d}t}=\frac{\partial X}{\partial t}+\vec u\cdot\vec\nabla X \to \frac{X}{\tau},
\end{equation}
for any quantity $X$. Another simplification is that the background turbulence forces the flow with the same time scale, so that the forcing term can be written $\vec u_0/\tau$.

In presence of a shear $\vec S=r\sin\theta\,\vec\nabla\Omega$, the flow is thus described by the continuity equation $\vec\nabla\cdot\vec u=\vec 0$ and the momentum and heat transport equations for adiabatic motions, where diffusion and viscosity are neglected as a first step, read:
\begin{align}
\frac{\vec u}{\tau} + 2\vec\Omega\wedge\vec u +(\vec S\cdot\vec u)\vec e_\varphi+\frac{\vec\nabla P}{\rho}-\vec g	&= \frac{\vec u^0}{\tau},	\\
\frac{s'}{\tau}																	&= -\vec u\cdot\vec\nabla\langle s\rangle.
\end{align}
Here, $s'$ denotes the specific entropy fluctuations and $\langle s\rangle$ the mean specific entropy. We also introduced the unit-vector basis $\left\{{\vec e}_{j}\right\}_{j=r,\theta,\varphi}$ in spherical coordinates.  Following \cite{Zahn1992}, we assume here a priori that $D_{\rm v}\!<\!\!<\!D_{\rm h}$ leading to a shellular differential rotation and corresponding shear which mainly depends on the radial coordinate. We thus consider here the case of a purely radial shear $\vec S=S\vec e_r$. This hypothesis will be verified in Appendix \ref{appendix:shellular}.

After combining the momentum and heat transport equations and eliminating the pressure fluctuations thanks to the continuity equation, one obtains in the Fourier space
\begin{equation}
\frac{\tilde{\vec u}}{\tau} + 2(\hat{\vec k}\cdot\vec\Omega)\hat{\vec k}\wedge\tilde{\vec u} +S\tilde u_r(\vec e_\varphi-\hat k_\varphi\hat{\vec k})+ \tau \tilde u_rN^2(\vec e_r - \hat k_r\hat{\vec k}) = \frac{\tilde{\vec u}^0}{\tau},
\end{equation}
where $\tilde{\vec u}$ is the Fourier transform of $\vec u$ and $\hat{\vec k}=\vec k/\|\vec k\|$ is the normalized wave vector. This can be written as a matrix equation
\begin{equation}
\mathcal{M}\cdot\tilde{\vec u} = \tilde{\vec u}^0,
\end{equation}
with
\begin{equation}
\mathcal{M}_{ij} = \delta_{ij} + \sigma\Omega^*\varepsilon_{ilj}\hat k_l +S^*(\delta_{i\varphi}-\hat k_\varphi\hat k_i)\delta_{jr} +{N^*}^2(\delta_{ir} - \hat k_r\hat k_i)\delta_{jr},
\end{equation}
where $\sigma=\hat{\vec k}\cdot\vec\Omega / \Omega$, $\Omega^* = 2\tau\Omega$, $S^*=\tau S$, $N^* = \tau N$, and $\delta$ and $\varepsilon$ are the usual Kronecker and Levi-Civita symbols respectively.
The matrix equation can be solved, using
\begin{eqnarray}
\mathcal{M}^{-1} = 
D^{-1}
\left\{\left[1+{N^*}^2(1-\mu^2)\right]\delta_{ij}+\sigma^2{\Omega^*}^2\hat k_i\hat k_j + \sigma\Omega^*\varepsilon_{ijl}\hat k_l\right.	\nonumber \\ 
+ \sigma\Omega^*{N^*}^2\varepsilon_{ilr}\mu\hat k_j\hat k_l + {N^*}^2\left(\mu\hat k_i - \delta_{ir}\right)\delta_{jr}	\nonumber \\
-S^*\mu\hat k_\varphi\delta_{i\varphi}\delta_{j\varphi}+S^*\varepsilon_{jl\varphi}\hat k_l\hat k_\varphi\delta_{i\theta} -S^*(1-\hat k_\varphi^2)\delta_{i\varphi}\delta_{jr}	\nonumber \\
\left.-\,\sigma\Omega^*S^*(1-\hat k_\varphi^2)\hat k_j\delta_{i\theta} -\sigma\Omega^*S^*\hat k_\theta\hat k_\varphi\hat k_j\delta_{i\varphi}\right\},
\end{eqnarray}
where
\begin{equation}
\label{eq:exprD}
D=1+{N^*}^2(1-\mu^2)+\sigma^2{\Omega^*}^2-S^*\mu\hat k_\varphi - \sigma\Omega^*S^*\hat k_\theta
\end{equation}
and $\mu=\hat k_r$. Some terms of $\mathcal{M}^{-1}$ vanish when applied to $\tilde{\vec u}^0$, because the latter also satisfies the continuity equation $\hat k_j\tilde u_j^0=0$. This finally leads to
\begin{equation}
\tilde{\vec u} = \mathcal{A}\cdot\tilde{\vec u}^0,
\end{equation}
with
\begin{equation}
\begin{aligned}
D\mathcal{A}_{ij} =\ &\left[1+{N^*}^2(1-\mu^2)\right]\delta_{ij} + {N^*}^2\left(\mu\hat k_i - \delta_{ir}\right)\delta_{jr}	+ \sigma\Omega^*\varepsilon_{ijl}\hat k_l\\
& -S^*\,\mu\hat k_\varphi\delta_{i\varphi}\delta_{j\varphi} + S^*\varepsilon_{jl\varphi}\hat k_l\hat k_\varphi\delta_{i\theta}-S^*(1-\hat k_\varphi^2)\delta_{i\varphi}\delta_{jr}\,.
\end{aligned}
\end{equation}

We introduce the spectral tensor $\tilde{\mathcal{Q}}^0$, which characterizes the properties of the background turbulence. It is defined by
\begin{equation}
\left\langle\tilde u^0_i(\vec k)\tilde u^0_j(\vec k')\right\rangle = \tilde{\mathcal{Q}}^0_{ij}\delta(\vec k+\vec k'),
\end{equation}
where $\langle\rangle$ denotes a statistical average. Then, it is possible to derive the spectral tensor of the generated turbulence $\tilde{\mathcal{Q}}$ by using the relation 
\begin{equation}
\label{eq:tensrel}
\tilde{\mathcal{Q}}_{ij} = \mathcal{A}_{im}\mathcal{A}_{jn}\tilde{\mathcal{Q}}^0_{mn}\,.
\end{equation}
In the case of purely horizontal random motions, which are an idealized view of turbulent displacements in a strongly stratified medium, we have
\begin{equation}
\tilde{\mathcal{Q}}^0_{ij}=\frac{3E(k)}{8\pi k^2}\left[(1-\mu^2)(\delta_{ij}-\hat k_i\hat k_j) - (\delta_{ir}-\mu\hat k_i)(\delta_{jr}-\mu\hat k_j)\right],
\end{equation}
where $E(k)$ is the kinetic energy spectrum function defined by
\begin{equation}
\left\langle (\vec u^0)^2\right\rangle = \int_0^\infty E(k){\rm d}k.
\end{equation}
The full expression for the tensor $\tilde{\mathcal{Q}}$ is too complex to be shown here. Instead, we focus on the components needed to determine the anisotropy of the transport using Eq.~\eqref{eq:ratio}. We have
\begin{align}
\tilde{\mathcal{Q}}_{rr} &= \frac{3E(k)}{8\pi k^2D^2}\sigma^2{\Omega^*}^2(1-\mu^2)^2,	\label{eq:Qrr}\\
\tilde{\mathcal{Q}}_{\theta\theta} &= \frac{3E(k)}{8\pi k^2D^2}\left\{\left[1+{N^*}^2(1-\mu^2) -S^*\mu \right]\hat k_\varphi - \sigma\Omega^*\mu\hat k_\theta\right\}^2,	\\
\tilde{\mathcal{Q}}_{\phi\phi} &= \frac{3E(k)}{8\pi k^2D^2}\left\{\left[1+{N^*}^2(1-\mu^2)-S^*\mu\hat k_\varphi\right]\hat k_\theta+\sigma\Omega^*\mu\hat k_\varphi\right\}^2.
\label{eq:Qpp}
\end{align}

Assuming that $N^*\gg1$ (i.e. $N\gg\tau^{-1}$) and $N^*\gg\Omega^*$ (i.e. $N\gg2\Omega$ as expected in the bulk of stellar radiation zones), these expressions can be simplified by retaining only the dominant terms. In this regime, Eq.~\eqref{eq:exprD} becomes
\begin{equation}
D\simeq {N^*}^2(1-\mu^2)
\end{equation}
and Eqs.~\eqref{eq:Qrr} to \eqref{eq:Qpp} yield
\begin{align}
\tilde{\mathcal{Q}}_{rr} &\simeq \frac{3E(k)}{8\pi k^2}\frac{\sigma^2{\Omega^*}^2}{{N^*}^4},	\label{eq:Qrr_bis}\\
\tilde{\mathcal{Q}}_{\theta\theta} &\simeq \frac{3E(k)}{8\pi k^2}\hat k_\varphi^2,	\\
\tilde{\mathcal{Q}}_{\varphi\varphi} &\simeq \frac{3E(k)}{8\pi k^2}\hat k_\theta^2,		\label{eq:Qpp_bis}
\end{align}
which show no explicit dependence on the shear number $S^*$. Therefore, we recover in the case of a radial differential rotation the expressions found by \cite{KB2012} in the case of solid-body rotation.\\

Turbulent velocity correlations, which allow to evaluate transport properties, are finally obtained thanks to the relation
\begin{equation}
\langle u_iu_j\rangle = \int\tilde{\mathcal{Q}}_{ij}{\rm d}\vec k.
\end{equation}
We define the two angles $\alpha$ and $\beta$ such that
\begin{align}
\hat k_r	&= \cos\alpha,	\\
\hat k_\theta	&= \sin\alpha\cos\beta,	\\
\hat k_\varphi	&= \sin\alpha\sin\beta.
\end{align}
The infinitesimal wave vector element can be written
\begin{equation}
{\rm d}\vec k = k^2 \sin\alpha\,{\rm d}k\,{\rm d}\alpha\,{\rm d}\beta,
\end{equation}
where $k$ is the norm of $\vec k$, and we now have
\begin{equation}
\sigma = \cos\theta\cos\alpha - \sin\theta\sin\alpha\cos\beta.
\end{equation}
Equations~\eqref{eq:Qrr_bis} to \eqref{eq:Qpp_bis} lead to
\begin{align}
\langle u_r^2\rangle &\simeq \langle (\vec u^0)^2\rangle\frac{{\Omega^*}^2}{2{N^*}^4},	\\
\langle u_\theta^2\rangle \simeq \langle u_\varphi^2\rangle &\simeq \frac{\langle (\vec u^0)^2\rangle}{2}.
\end{align}
In the previously introduced notations, one has $u_\perp^2=\langle u_\theta^2\rangle + \langle u_\varphi^2\rangle$ and $u_\parallel^2=\langle u_r^2\rangle$.
Using Eq.~\eqref{eq:ratio}, we finally obtain
\begin{equation}
\label{eq:anisoKB}
\frac{D_{\rm v}}{D_{\rm h}}=\frac{1}{2}\frac{(2\Omega)^2}{N^4\tau^2}.
\end{equation}
As expected, the strength of the horizontal turbulent transport relatively to the vertical one is thus increasing with stratification (as $N^4$). On the other hand, it decreases with rotation (as $\Omega^{-2}$) since rotation is limiting horizontal turbulent motions through the Coriolis acceleration.

This result constitutes a new key to understand the anisotropic turbulent transport in stellar radiation zones. If we are 
able to compute the turbulent transport in the vertical (horizontal) direction induced by a given instability, it allows us 
to deduce the induced transport in the horizontal (vertical) one as a function of stratification and rotation. This strongly improves the current state of the art with respect to prescriptions previously proposed for $D_{\rm h}\equiv D_{\rm h,h}$ \citep{Zahn1992,Maeder2003,MPZ2004} for the turbulent transport induced by the instability of the horizontal shear, that were based on phenomenological laws and which did not take the simultaneous action of the stratification and the Coriolis acceleration into account.

However, the time $\tau$ that characterizes the turbulence is still a free parameter in Eq. (\ref{eq:anisoKB}), and should be specified. The only constraint proposed by \cite {KB2012} is that it should be much larger than the time characterizing the buoyancy $\tau_{N}=1/N$.

\subsubsection{Characterization of the turbulent timescale}
\label{subsec:rotation}
We propose here three physically motivated possibilities to estimate the turbulent timescale: (i) the time characterizing the radial shear, $\tau=1/S$, where $S=r\sin\theta\partial_r\Omega$; (ii) a time characterizing the Coriolis acceleration of a rotating radially sheared flow, $\tau=1/(2\Omega+S)$; and (iii) the time associated to the epicyclic frequency $N_\Omega=\sqrt{2\Omega(2\Omega+S\sin\theta)}$, which is one of the frequencies characterizing differentially rotating flows and their stability, $\tau=1/N_\Omega$.\\

First, we consider the case of unstable non-rotating vertically sheared flows. The results of numerical simulations of shear instability without rotation \citep{Pratetal2014} suggest that the typical turbulent timescale, which corresponds to the turnover time of large eddies, is of the order of $1/S$ in the non-rotating strongly stratified regime. Therefore, $\tau=1/S$ seems to be a reasonable choice when the dynamics is dominated by the shear.  

Moreover, this choice allows one to recover the model of vertical turbulent transport by \citet{Zahn1992} using arguments on time scales. Indeed, the fact that turbulence can be sustained only if the dynamical timescale $\tau$ is smaller than the timescale of the process that inhibits turbulence in the vertical direction, here the stratification modified by thermal diffusion, can be written
\begin{equation}
\tau < \frac{K}{N^2 l_{\parallel}^2},
\end{equation}
where $K$ is the thermal diffusivity and $l_{\parallel}$ the typical turbulent length scale along the stratification direction. The scales that induce the most important transport are the largest unstable scales given by
\begin{equation}
l_{\parallel}^2 = \frac{K}{N^2 \tau}.
\end{equation}
The turbulent transport along the vertical direction is then of the order of $S l_{\parallel}^2$, which yields
\begin{equation}
D_{\rm v,v} \sim \frac{K S}{N^2\tau}.
\end{equation}
Choosing $\tau = S^{-1}$ finally implies
\begin{equation}
D_{\rm v,v} \sim \frac{K S^2}{N^2},
\end{equation}
which is precisely the model by \citet{Zahn1992}. We recall that this turbulent transport occurs if and only if the turbulent Reynolds number (i.e. the ratio between the viscous time and the dynamical time of turbulence) $R_{\rm e}=\left(u_{\parallel}\,l_{\parallel}\right)/\nu$, where $\nu$ is the viscosity of the fluid, is larger than a critical value $R_{\rm e;c}$ \citep{Zahn1992,TZ1997,Pratetal2016}. In the STAREVOL code, we chose to set $R_{\rm e;c} = 10$ that corresponds to a value of $D_{\rm v,v}$ of the order of ten times the molecular viscosity.
\\

When the dynamics is no longer dominated by the shear because of rotation,  $\tau=1/(2\Omega+S)$ becomes a plausible choice. Indeed, the frequency $2\Omega + S$ characterizes the Coriolis acceleration along the azimuthal direction in a rotating sheared flow \citep[e.g.][]{Mathis2009}. Moreover, it allows one to recover $\tau=1/S$ in the asymptotic limit where $\Omega \ll S$.
In contrast, when the shear is negligible, $\tau\simeq 1/(2\Omega)$, which is the characteristic time scale of inertial waves. This can be expected since inertial waves are known to structure turbulence in rapidly rotating flows because of their non-linear interactions \citep[e.g.][]{Senetal2012}. 

Another frequency that characterizes  differentially rotating fluids and their stability, the epicyclic frequency \citep[e.g.][]{PringleKing2007}
\begin{equation}
N_{\Omega}^{2}=\frac{1}{{\varpi}^3}\frac{{\rm d}}{{\rm d}\varpi}\left(\varpi^4\Omega^2\right)=2\Omega\left(2\Omega+\varpi\frac{{\rm d}\Omega}{{\rm d}{\varpi}}\right),
\end{equation}
with $\varpi=r\sin\theta$, can also be considered with the corresponding characteristic time $\tau=1/N_\Omega$. In the limit where the shear is negligible, one obtains $\tau\simeq 1/(2\Omega)$, as with $\tau=1/\left(2\Omega+S\right)$. In the opposite limit where rotation is negligible, $\tau\sim\sqrt{2\Omega S\sin\theta}$, which is not compatible with the two other choices ($\tau=1/S,1/\left(2\Omega+S\right)$), since $\tau$ still depends on $\Omega$. 

We should here point out that one will be able to make a choice between $\tau=1/\left(2\Omega+S\right)$ and $1/N_{\Omega}$ only when  direct numerical simulations of the  vertical shear instability in rotating stably stratified fluids will be available for stellar regimes. This is the reason why we will here study and discuss both cases. Another important point to keep in mind when choosing one of these characteristic times is to verify that the rotation profile is not subject to the Rayleigh-Taylor instability for which $N_\Omega^2<0$. In practice, this is generally not a problem during the main-sequence since $N_\Omega^2>0$ at this phase. However, for negative values of $N_\Omega^2$, one would need to model the effect of the Rayleigh-Taylor instability itself and the turbulent transport it induces \citep[][]{Maederetal2013}.

Finally, if one considers the choice $\tau=1/N$, even though \citet{KB2012} assumed it was not the case, one obtains an anisotropy which is consistent with the work by \citet{BC2001}. However, $1/N$ characterizes the stratification that inhibits the turbulence while the nonlinear timescale should rather be given by the process that drives the turbulence, i.e. the shear. Therefore, we rule out this possibility in the following study.

\subsubsection{Horizontal turbulent transport}
\label{Sect:Prescrp_Dh}

Using Eq. (\ref{eq:anisoKB}), it is finally possible to derive $D_{\rm h}$ when knowing $D_{\rm v}$ (or vice-versa):
\begin{equation}
\frac{D_{\rm h}}{D_{\rm v}}=\frac{1}{2}\frac{N^4}{\Omega^2}{\tau^2}
\label{Anis3}
\end{equation}
for a given choice of turbulent source and characteristic time $\tau$.\\
In stellar radiation zones, the main important source of turbulence in the vertical direction is the secular vertical shear instability \citep{Zahn1992,TZ1997}. This instability induces 3D turbulent motions both in the vertical and in the horizontal directions \citep[e.g.][]{Pratetal2013,Garaud2016}. The horizontal turbulent diffusion coefficient computed using Eq. (\ref{Anis3}), when assuming that $D_{\rm v}$ is associated with this secular radial shear instability (i.e. $D_{\rm v}\equiv D_{\rm v,v}$ and thus $D_{\rm h}\equiv D_{\rm h,v}$) and non-zero (for $R_{\rm e}>R_{\rm e;c}$), thus corresponds to the horizontal turbulent transport induced by 3D anisotropic turbulent motions this vertical instability sustains because of the stable stratification and rotation. 

It is important to point out that if the horizontal shear 
becomes large enough to be simultaneously unstable, it will induce a horizontal turbulent transport with the corresponding turbulent diffusion coefficient \citep[$D_{\rm h,h}$, see][]{MPZ2004} as well as a vertical transport (and corresponding vertical turbulent diffusion coefficient {\bf $D_{\rm v,h}$}) because of the 3D anisotropic turbulent motions this horizontal instability would sustain.  This was already pointed out by \cite{Zahn1992} (in his section \S 2.4.2), even if  it has never been considered in stellar evolution studies.\\

In this work, we choose to focus on the first of these two cases. To describe the turbulent transport along the radial direction due to the secular radial shear instability, we adopt the model derived by \cite{Zahn1992}:
\begin{equation}
D_{\rm v,v}=\frac{Ri_{\rm c}}{3} K \left(\frac{r\sin\theta\,\partial_{r}\Omega}{N}\right)^{2}\quad\hbox{if}\quad R_{\rm e}>R_{\rm e;c},
\label{eq:Dv}
\end{equation}
which has been recently examined and validated by independent high-resolution Cartesian non-linear numerical simulations \citep{Pratetal2013,Pratetal2016,Garaud2016}. The horizontal turbulent diffusion coefficient associated with the horizontal turbulent transport induced by 3D motions that the vertical instability sustains when $R_{\rm e}>R_{\rm e;c}$, $D_{\rm h,v}$, is derived using Eq. (\ref{Anis3}).\\

The first possible choice for $\tau$, namely, $\tau,= S^{-1}$ leads to
\begin{equation}
D_{\rm h,v}=\begin{cases}\displaystyle{\frac{2Ri_{\rm c}}{3}\left(\frac{N}{2\Omega}\right)^2 K}\quad\hbox{if}\quad R_{\rm e}>R_{\rm e;c}\\
0\quad\hbox{if}\quad R_{\rm e}<R_{\rm e;c}.
\end{cases}
\label{eq:Dh}
\end{equation}
In this case, we thus obtain a discontinuous variation for $D_{\rm h,v}$ if the Reynolds criterion is not met and the vertical shear-induced turbulence vanishes. This may be regularized once a prescription for $D_{\rm v,v}$ that will take into account the action of the Coriolis acceleration will be derived. Since we expect the Coriolis acceleration to stabilize the flow, the transport predicted by Eq. (\ref{eq:Dh}) can be considered as an upper limit of the actual transport efficiency. 

We then examine the two choices for characteristic turbulent timescales that are already including rotation. First, assuming a  shellular rotation (see Appendix \ref{appendix:shellular}), the expression
\begin{equation}
\tau=\left(2\Omega+S\right)^{-1}=\left(2\Omega+r\sin\theta\partial_{r}\Omega\right)^{-1}
\end{equation}
leads to
\begin{equation}
{D_{\rm h,v}}=\begin{cases}\displaystyle{\frac{2Ri_{\rm c}}{3}\left(\frac{N}{2\Omega}\right)^{2}K\frac{\sin^{2}\theta}{\left(X^{-1}+\sin\theta\right)^2}}\quad\hbox{if}\quad R_{\rm e}>R_{\rm e;c}\\
0\quad\hbox{if}\quad R_{\rm e}<R_{\rm e;c}
\end{cases},
\label{eq:Dh2_primitive}
\end{equation}
where $X=r\partial_{r}\Omega/(2\Omega)$. 
For the implementation in  1D stellar evolution codes,  we take a latitudinal average of this expression
\begin{equation}
{D_{\rm h,v}}=\begin{cases}\displaystyle{\frac{2Ri_{\rm c}}{3}\left(\frac{N}{2\Omega}\right)^{2}K\left<\frac{\sin^{2}\theta}{\left(X^{-1}+\sin\theta\right)^2}\right>_{\theta}}\quad\hbox{if}\quad R_{\rm e}>R_{\rm e;c}\\
0\quad\hbox{if}\quad R_{\rm e}<R_{\rm e;c}
\end{cases}
,
\label{eq:Dh2}
\end{equation}
where  $\left<...\right>_{\theta}=\int_{0}^{\pi}...{\sin\theta}\,{\rm d}\theta/\int_{0}^{\pi}{\sin\theta}\,{\rm d}\theta$. \\ 
Finally, considering the epicyclic  characteristic time
\begin{equation}
\tau=\frac{1}{N{_\Omega}}=\left[2\Omega\left(2\Omega+r\partial_{r}\Omega\sin^{2}\theta\right)\right]^{-1/2},
\end{equation}
we obtain
\begin{equation}
{D_{\rm h,v}}=\begin{cases}\displaystyle{\frac{2Ri_{\rm c}}{3}\left(\frac{N}{2\Omega}\right)^{2}K\left< \frac{X\sin^2\theta}{X^{-1}+\sin^2\theta}\right>_{\theta}}\quad\hbox{if}\quad R_{\rm e}>R_{\rm e;c}\\
0\quad\hbox{if}\quad R_{\rm e}<R_{\rm e;c}
\end{cases}.
\label{eq:Dh3}
\end{equation}
The horizontal averages in Eqs. (\ref{eq:Dh2}) and (\ref{eq:Dh3}) are analytically computed in the Appendix \ref{appendix:HA} to allow a robust numerical implementation.

Interestingly, all the choices for $\tau$ lead to the same dependence of $D_{\rm h,v}$ on the ratio $N/(2\Omega)$. Including rotation in the prescription for $\tau$ also allows us to regularize the behavior of $D_{\rm h,v}$ for a vanishing vertical shear-induced turbulence. This is a significant improvement towards a consistent description of anisotropic turbulence in stellar radiation zones. As already pointed out above, the next step will require a model for $D_{\rm v,v}$ that takes the action of the Coriolis acceleration into account. This should be achieved with direct numerical simulations of the  vertical shear instability in rotating stably stratified fluids in stellar regimes, which are beyond the scope of this work, and which will allow us to test our analytical prescriptions. In this framework, it will be important to take also the action of the chemical stratification (through the $\mu$-gradients) and of the viscosity into account as in \cite{Pratetal2016}. Finally, it will be necessary to take into account the possible feed-back of the horizontal turbulence on the vertical turbulent transport \citep{TZ1997}. 

In the following section, we investigate the impact of the different expressions we derived for $D_{\rm h,v}$ on the transport of angular momentum along the evolution of low-mass main sequence (MS), subgiant and red giant stars. 


\section{Application to stellar interiors and evolution}
\label{sec:appli}

\subsection{Method}
\label{methodcalcul}
We implement  Eqs.~(\ref{eq:Dv}), (\ref{eq:Dh}), (\ref{eq:Dh2}) and (\ref{eq:Dh3}) in the stellar evolution code STAREVOL as an additional source of turbulence in the horizontal direction. The total horizontal turbulent diffusion coefficient D$_{\rm h}$ is then defined as the sum of D$_{\rm h,v}$ as given by these equations (we test the different expressions for $\tau$) and of D$_{\rm h,h}$ from \citet[][Eq.~19]{MPZ2004}. The treatment of rotation-induced processes in STAREVOL is described in detail e.g. in \citet{Decressinetal2009} and \citet{Amardetal2016}, and we refer to this latest paper for the information on the basic input physics (i.e., equation of state, nuclear reactions, opacities, etc.) used here 
(see also \citealt{Siessetal2000,Palaciosetal2003,Lagardeetal2012}). 
The angular velocity profile in the radiative interior derives from the redistribution of angular momentum by shear-induced turbulence and meridional flows, which are treated using the formalism derived by \cite{Zahn1992}, \cite{MZ1998}, and \cite{MZ2004}; uniform rotation is assumed in convective regions. Angular momentum losses at the stellar surface due to pressure-driven magnetized stellar winds are taken into account using the prescription by \citet{Matt15}.
Convective regions are modeled according to the mixing length theory. The adopted mixing-length parameter is defined by the calibration of a standard solar model and is fixed to $\alpha_{\rm MLT}=1.973$. We adopt $Z = 0.013446$ as the initial metallicity that corresponds to the solar metallicity according to \citet{AGSS09}.

We explore the impact of our new 
modeling of the shear-induced horizontal turbulence on the transport of angular momentum in low-mass stars along the MS and the red giant branch. We start with the case of a 1~M$_\odot$, Z$_\odot$  model for which we have constraints both on the internal rotation rate at the age of the Sun and on the evolution of the surface rotation with time (\S~\ref{stellarmodels}). We then move to the case of a 1.25~M$_\odot$, Z$_\odot$ model which internal and surface rotation states are constrained by asteroseismic data from mixed modes of subgiant and giant stars (\S~\ref{1.25model}).

\subsection{The case of a 1.0 M$_\odot$, Z$_\odot$ main sequence star}
\label{stellarmodels}

We focus on a 1.0~M$_\odot$ star at solar metallicity and 
present a comparative analysis of the evolution of the different diffusion coefficients and  meaningful quantities as a function of time along the MS evolution of this low-mass star. For this, we compare several models computed from the pre-main sequence (PMS) up to 9 Gyr with the different expressions for $\tau$ (i.e., with the different expressions for the horizontal eddy diffusivity $D_{\rm h,v}$ given in Eqs.~(\ref{eq:Dh}), (\ref{eq:Dh2}), and (\ref{eq:Dh3}), added to the horizontal turbulent diffusivity $D_{\rm h,h}$ derived in \citealt{MPZ2004}), keeping all the other ingredients unchanged. These models are also compared with a model that ignores $D_{\rm h,v}$ and includes only the prescription for $D_{\rm h,h}$ given by \citet{MPZ2004}.

\subsubsection{Model setup}
\label{modelsetup}
We use similar assumptions for the initial rotation rate and prescription for magnetic braking as in \citet{Amardetal2016}.
A phase of disc coupling, during which the surface rotation rate is maintained to its original value of $\Omega_{\rm ini} = 1.6 \times 10^{-5}$ s$^{-1}$ (corresponding to an initial rotation period of 4.55 days), is accounted for during the first 5 Myr of the evolution on the PMS. This corresponds to a median rotator, as observed in statistical samples of PMS stars in young clusters (see \citealt{Amardetal2016} and \citealt{GB2015} for details). Four median rotator models are thus computed, three with the different expressions for $D_{\rm h,v}$ added to $D_{\rm h,h}$ from \citet{MPZ2004}, and the last one with only $D_{\rm h,h}$.
For the case where $D_{\rm h,v}$ is computed with $\tau=1/S$ (Eq.~\ref{eq:Dh}), we also compute two models with different initial velocities to investigate the effect of the global velocity on the turbulent transport; these fast and slow rotators start with an initial period of 1.4 and 9.0 days, respectively. For all the models angular momentum losses at the stellar surface due to pressure-driven magnetized stellar winds are taken into account using the prescription by \citet{Matt15} with $K_{\rm wind}=7\times10^{30}$ to fit the solar rotation rate at 4.57 Gyr, $m=0.22$ as expected from a purely dipolar magnetic field and the parameter $p=2.4$ to fit the surface rotation rates in open clusters at best. 
We then cover the statistical distributions of  rotation  periods in  open clusters and  associations from ~1 Myr to 2.5~Gyr \citep{GB2013,GB2015,Amardetal2016}.
We summarize in Table~\ref{tab:1Msunmodels} the specifications of the different 1~M$_\odot$, Z$_\odot$ models that we compare below.

\begin{table}[h]
\begin{center}
\caption{Specifications of the different 1~M$_\odot$, Z$_\odot$ models.
In all cases $D_{\rm h,h}$ is included using \citet{MPZ2004} prescription and added to $D_{\rm h,v}$ given in column 3, except in case D where we use $D_{\rm h,h}$ only. The last column refers to the colors used for the different models in Figs.~\ref{Fig:Hor_diag}, \ref{Fig:Anis}, \ref{Fig:Deff}, \ref{Fig:Omega_prof}, and \ref{Fig:Diag_Bouvier}.}
\begin{tabular}{ c c c c c }
\hline
\hline
Model & $\tau$ & D$_{\rm h,v}$ & rotation & color \\
\hline
A1 & $1/S$ & Eq.~\ref{eq:Dh} & median  & green \\
A2 & $1/S$ & Eq.~\ref{eq:Dh} & fast  & light green \\
A3 & $1/S$ & Eq.~\ref{eq:Dh} & slow & dark green \\
B & $1/(2\Omega+S)$ & Eq.~\ref{eq:Dh2} & median & blue \\
C & $1/N_\Omega$ & Eq.~\ref{eq:Dh3} & median & magenta \\
D & -- & -- & median & black \\
\hline
\end{tabular}
\label{tab:1Msunmodels}
\end{center}
\end{table}

\begin{figure*}[!ht]
\centering
\includegraphics[width=0.825\textwidth]{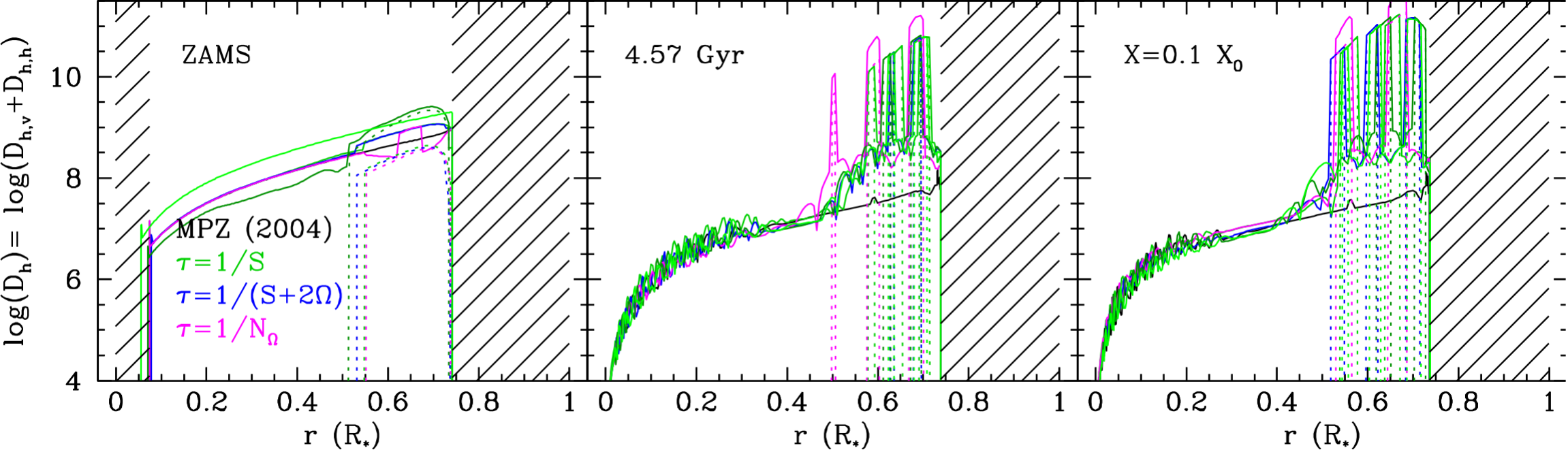}
\hspace*{0.2cm}
\includegraphics[width=0.815\textwidth]{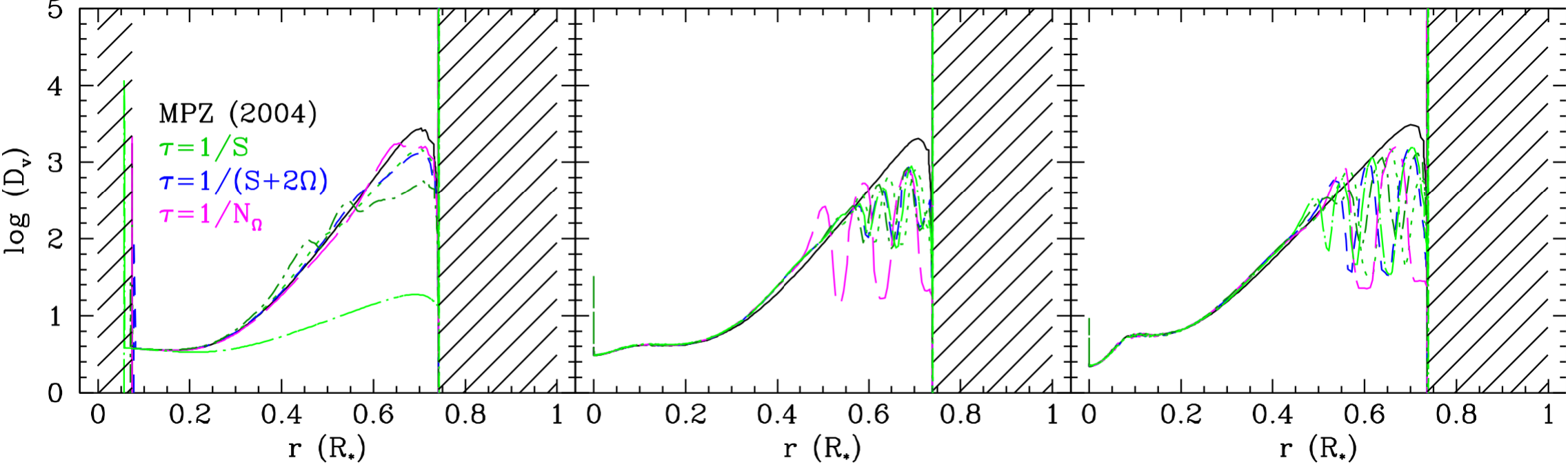}
\caption{{\it Top:} Total horizontal turbulent diffusivity \Dh{} (full line) and its $D_{\rm h,v}$ component (when active if $R_{\rm e}>R_{\rm e;c}$; dotted lines) for the 1.0 \Ms{}, Z$_\odot$  models described in Tab.~\ref{tab:1Msunmodels}
at three evolutionary stages (from left to right: ZAMS, age of the Sun, and when the central H mass fraction is reduced to 10$\%$ of its original value).  
The hatched areas correspond to the convective regions. 
{\it Bottom:} Vertical turbulent diffusion coefficient in the same models.
}
\label{Fig:Hor_diag}
\end{figure*}

\subsubsection{Comparing different prescriptions for the horizontal turbulent diffusivity}
\label{CompDh}

In Fig.~\ref{Fig:Hor_diag}, we compare the profiles of the total horizontal diffusivity $D_{\rm h}$ and of the vertical diffusivity $D_{\rm v,v}$ within all the 1.0 \Ms{} , Z$_\odot$  models described in Table~\ref{tab:1Msunmodels}
at three different epochs on the MS (Zero Age Main Sequence, age of the Sun, and when the central hydrogen mass fraction is reduced to 10$\%$ of its original value).
We note first that below $\sim 0.5 R_\star$, where $R_\star$ is the radius of the star, the novel expressions for the horizontal turbulent diffusion ($D_{\rm h,v}$) lead to values for the total $D_{\rm h}=D_{\rm h,v}+D_{\rm h,h}$ that are very close to the ones given by the expression of $D_{\rm h,h}$ by \cite{MPZ2004} alone. 
The main reason for this behavior is that the differential rotation needs to be high enough to trigger on the vertical shear instability (i.e. $R_{\rm e}>R_{\rm e;c}$), and this is not the case in the central part of the star for most of the MS. Therefore, the horizontal component of $D_{\rm h}$ ($D_{\rm h,h}$) given by \cite{MPZ2004} dominates the transport of angular momentum along the isobars, all along the MS in the deep interior. This can be seen clearly by looking at the behavior of  $D_{\rm h,v}$, which becomes important only in the outer part of the radiative region above $\sim 0.5 R_\star$ where differential rotation is stronger (and able to trigger the vertical shear instability), as imposed by the extraction of angular momentum by the magnetized wind. At the age of the Sun and up to the turn-off, $D_{\rm h,v}$ can be several orders of magnitude higher than $D_{\rm h,h}$ in these outer layers below the convective envelope.

\begin{figure*}
\centering
\includegraphics[width=0.8\textwidth]{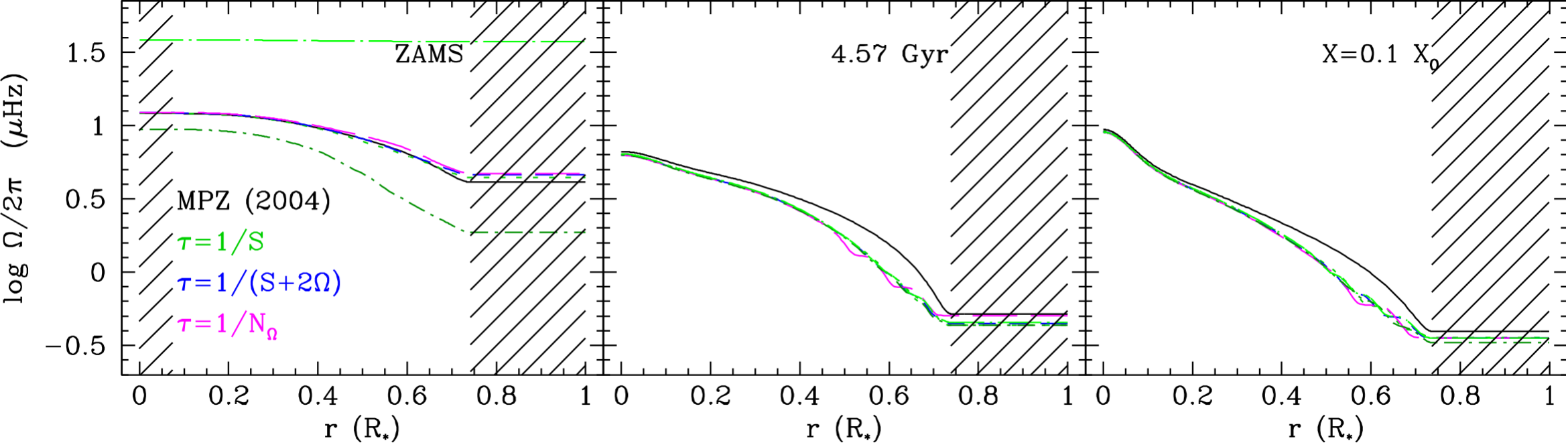}
\hspace*{0.1cm}
\includegraphics[width=0.785\textwidth]{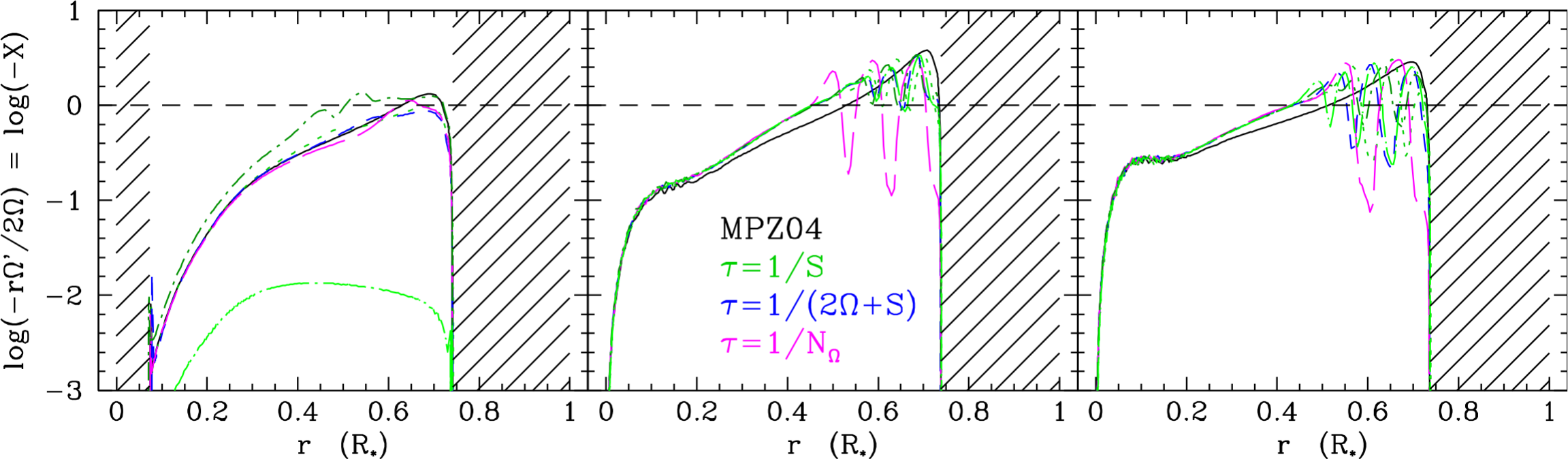}
\caption{{\it Top:} Angular velocity profile represented at the same evolutionary stages and with the same color code as in Fig~\ref{Fig:Hor_diag} {\it Bottom:} Ratio between the differential rotation and the rotation rate, equivalent to the dimensionless parameter X in Eqs.~\eqref{eq:Dh}, \eqref{eq:Dh2_primitive}, and \eqref{eq:Dh3}.}
\label{Fig:Omega_prof}
\end{figure*}

Figure~\ref{Fig:Hor_diag} also shows the profiles of the vertical turbulent diffusivity $D_{\rm v}=D_{\rm v,v}$ at the same evolutionary stages for all the prescriptions. 
The strong coupling at the ZAMS obtained in the fast rotator (light green curve, model A2) is due to a very efficient transport by the meridional circulation, and results in a weak vertical turbulent transport as can be seen in the corresponding bottom left panel.
As the models evolve on the MS a flattening of the \Dv{} profile appears below the convective envelope (above $r \approx 0.5 R_\star$) for models A1, A2, A3, B and C. The vertical shear-induced turbulent transport seems to be weaker in average around $\log(D_{\rm v})\approx 2.5$ in this region when the radial differential rotation is accounted in addition to the horizontal shear as a source for the horizontal turbulent transport (i.e. when adding $D_{\rm h,v}$ to the $D_{\rm h,h}$ of \citealt{MPZ2004}). The radial variations of \Dv{} can also be understood by the radial differential rotation obtained for $r>0.5 R_\star$ reported in Fig. \ref{Fig:Omega_prof}.

\begin{figure*}[th]
\begin{center}
\includegraphics[width=0.8\textwidth]{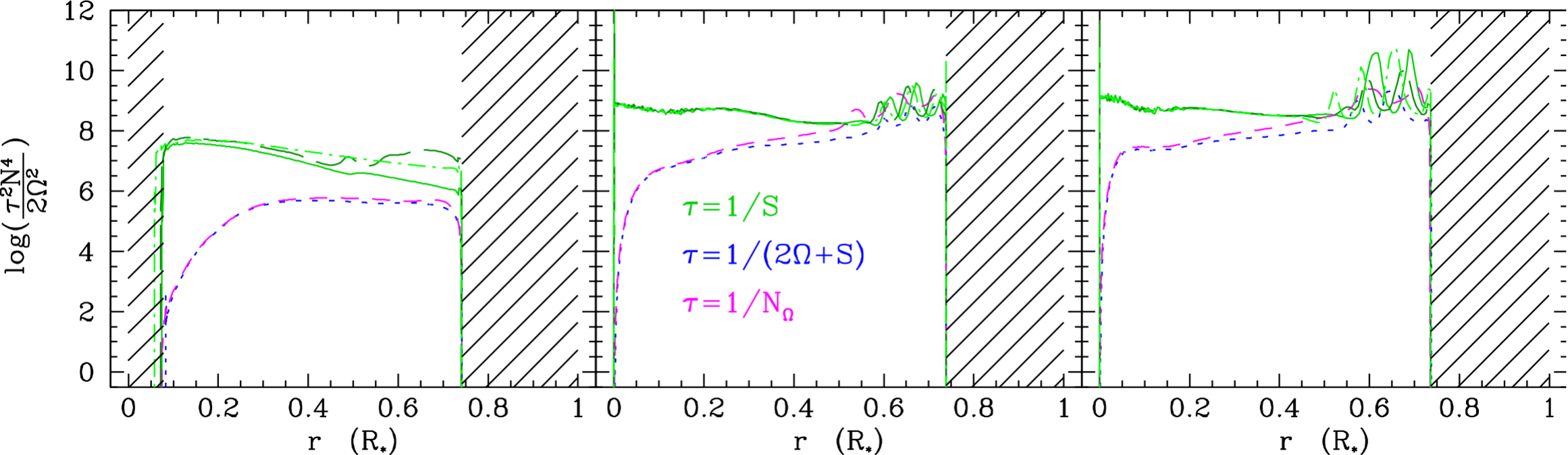}
\includegraphics[width=0.8\textwidth]{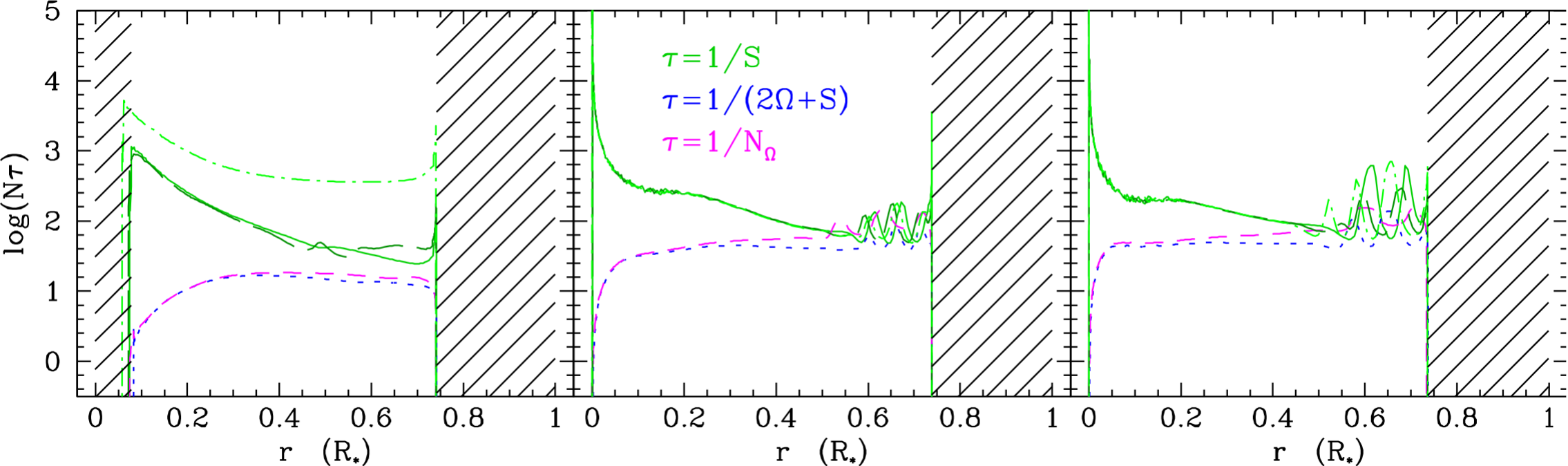}
\end{center}
\caption{{\it Top:} Anisotropic ratio given by Eq.~(\ref{Anis3}) for the different choices of $\tau$. It represents the potential anisotropy of turbulence sustained by the vertical shear instability in the stably stratified differentially rotating medium. It is thus equivalent to the ratio between the horizontal and vertical turbulent diffusivities ($D_{\rm h,v}/D_{\rm v,v}$) when the latter is active. {\it Bottom:} product of the thermal part of the Brunt-Väisälä frequency ($N$) by the computed turbulent characteristic time ($\tau$). Both are shown at the same evolutionary stages and with the same color code as in Fig~\ref{Fig:Hor_diag}.}
\label{Fig:Anis}
\end{figure*}

On the upper panel of Fig.~\ref{Fig:Anis} we show the anisotropy ratio of the turbulence given by Eq.~(\ref{Anis3}). Note that  this representation actually corresponds to the {\it potential} anisotropy of turbulence since we plot it even in the vertical shear-stable regions. 
We can distinguish two behaviors with the new prescriptions : the case of models B and C where the turbulent time-scale depends on the rotation rate on one hand, and the case of models A where such a dependence is not present ($\tau=1/S$) on the other hand. While both groups show an anisotropy ratio that increases moderately during the evolution, in the first group of models the anisotropy is relatively low in the central region due to the higher angular velocity there, while in the second case the anisotropy is strong in the central region.
Quantitatively during the MS, the region below the envelope shows an anisotropy ratio that is about $10^9$. This means that whenever the vertical shear is high enough to overcome the Reynolds criterion (i.e. $R_{\rm e}>R_{\rm e;c}$) below the convective envelope for $r>0.5R_\star$, the total horizontal diffusion coefficient $\log ({\rm D}_{\rm h})$ rises from $7.5$ up to $11.5$ $( \simeq 10^9 \times {\rm D}_{\rm v} )$ that leads to a stronger meridional circulation and transport of angular momentum as explained in Appendix \ref{appendix:TL}.

On the lower panel of Fig.~\ref{Fig:Anis} we show the product of the Brunt-V\"ais\"al\"a frequency $N$ (its thermal part given in Eq.\ref{BVf}) by the turbulent characteristic timescale $\tau$ for models A, B and C at the same evolution points on the MS.
As initially assumed by \citet{KB2012}, this quantity turns out to be greater than unity all along the MS evolution. 
It stays rather constant during this latter and shows very similar values in models B and C. Indeed, rotation tends to compensate for the disappearance of the shear, and even leads to a decrease of the turbulent time-scale, thus limiting the increase of the product $N\tau$. For models A though, the product $N\tau$ is high in the central region where the shear is weaker at early stages (see below, Fig.~\ref{Fig:Omega_prof}) and the Brunt-V\"ais\"al\"a frequency is important. 

\begin{figure*}
\centering
\includegraphics[width=0.8\textwidth]{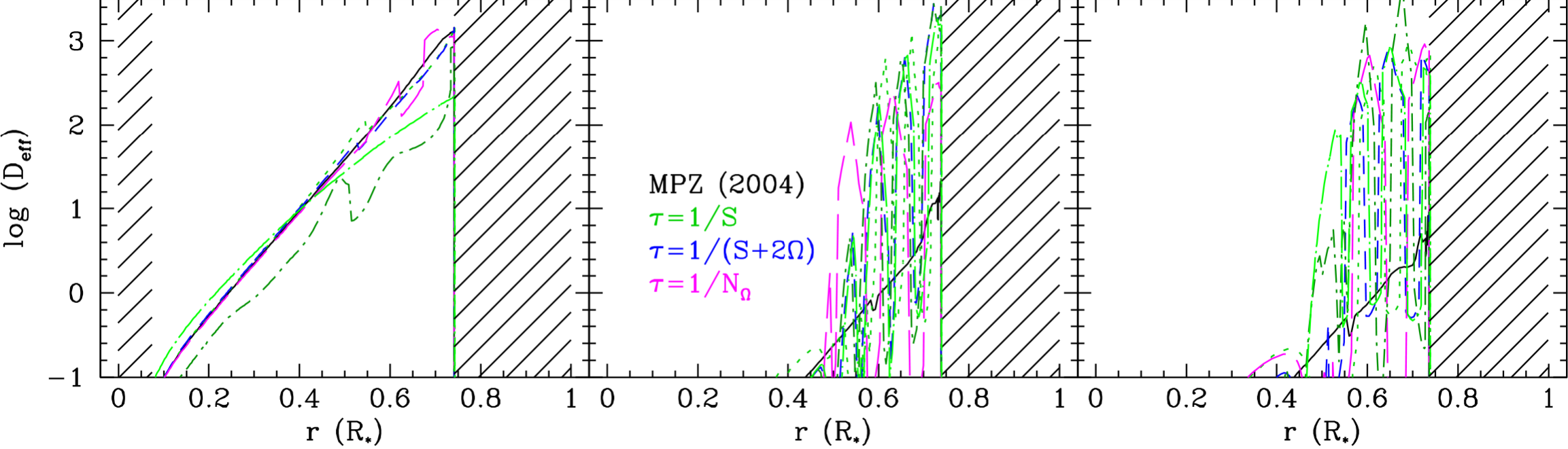}
\caption{Effective diffusivity \Deff in the vertical direction induced by the advective transport at the same evolutionary stages and with the same color code as in Fig~\ref{Fig:Hor_diag}.}
\label{Fig:Deff}
\end{figure*}

Finally, we show the effective diffusivity (Fig.~\ref{Fig:Deff}) that is also affected by the change of horizontal turbulent diffusion coefficient and the potentially enhanced meridional circulation (see Appendix \ref{appendix:TL}), consistently with what was already pointed out in \citet{MPZ2007}, the erratum associated with \citet{MPZ2004}. By definition, $D_{\rm eff} = \displaystyle{\frac{1}{30} \frac{(rU_{2})^2}{D_{\rm h}}}$, where $U_{2}\left(r\right)$ is the radial function of the expansion of the radial component of the meridional circulation on Legendre polynomials \citep{ChaboyerZahn1992}. In the case of a strong horizontal turbulence, one can see in \cite{Decressinetal2009} (we refer the reader to Eqs. 32 and 40 in this article) that entropy advection by the meridional circulation balances the horizontal turbulent heat diffusion. Then, we have $U_2\propto D_{\rm h}$ and $D_{\rm eff}\propto D_{\rm h}$. This explains both the amplitude and the radial variations of $D_{\rm eff}$ observed for $r>0.5R_\star$ in models A, B, and C.


\subsubsection{Impact on the evolution of the internal and surface rotation}
\label{subsec:RotEvol}


\begin{figure}[!ht]
\includegraphics[width=0.48\textwidth]{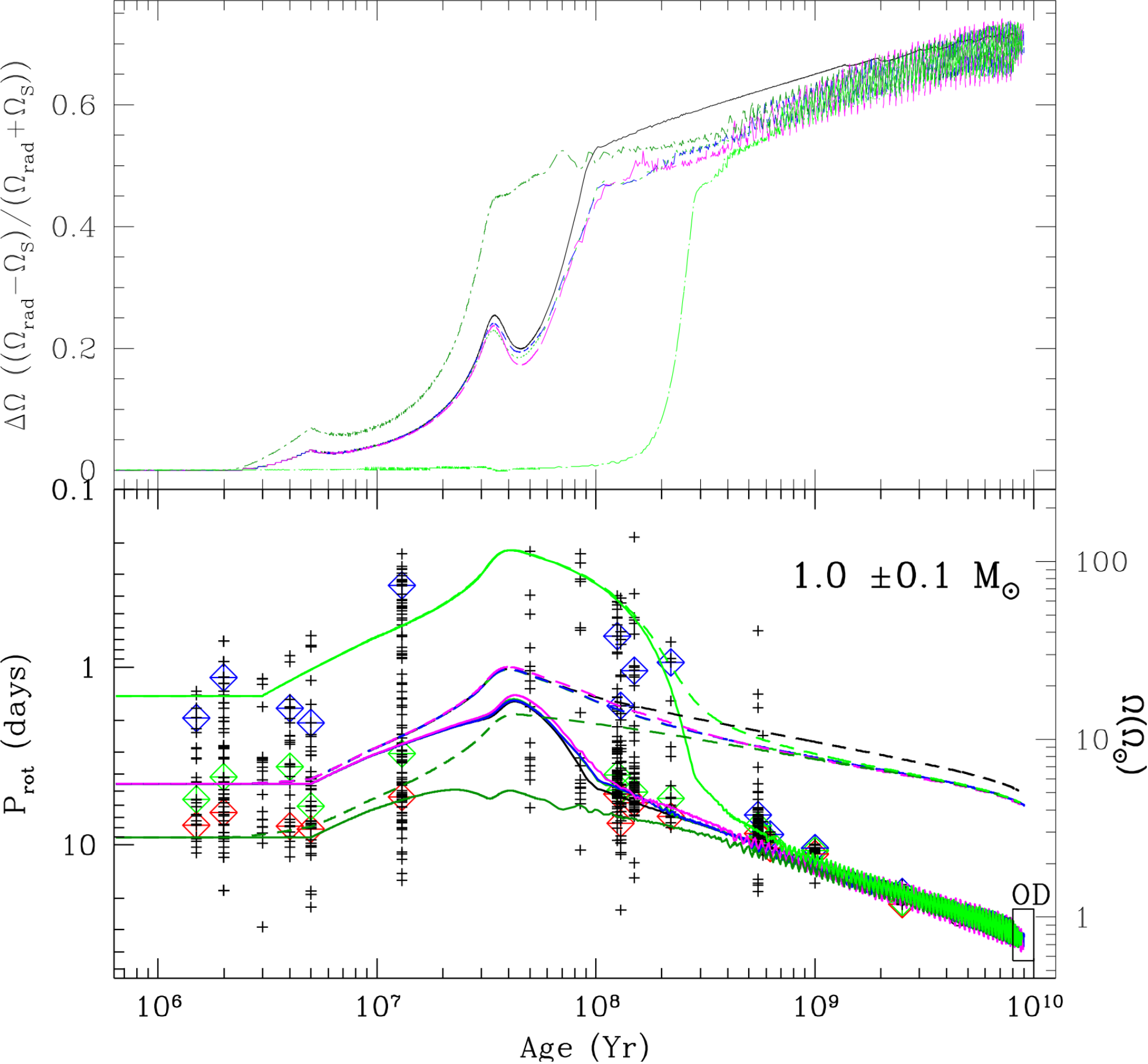}
\caption{{\it Bottom} Evolution of the angular velocity as a function of time for models computed with the prescriptions for \Dh{} from \cite{MPZ2004}, \cite{Zahn1992}, and that of the present paper with the three different characteristic time scales for turbulence (black, red, green ($\tau=1/S$), blue ($\tau=1/(2\Omega+S)$) and magenta ($\tau=1/N_\Omega$), respectively). Scaled on the left is the rotational period in days and on the right the angular velocity in solar units with $\Omega_\odot = 2.86.10^{-6}\ {\rm s}^{-1}$. Each cross represents a star belonging to an open cluster which age and mass (between 1.4 and 1.6\Ms) have been taken from literature. All observational data are taken from \cite{GB2015} and \cite{Bouvier2014} and references therein. The solid lines show the evolution of the surface rotation, and dashed lines the evolution of the mean radiative core rotation rate. {\it Top} Evolution of the normalised core to surface differential rotation rate for the six 1.0~M$_\odot$, Z$_\odot$ models as color-coded in Fig.~\ref{Fig:Hor_diag}.}
\label{Fig:Diag_Bouvier}
\end{figure}

We now discuss the impact of introducing the new formalism for \Dh{} on the evolution of internal rotation and on the surface rotation rates of the 1.0~M$_\odot$, Z$_\odot$ models. 
First, we verify and confirm in Appendix \ref{appendix:shellular} that the hypothesis of a shellular rotation are respected, i.e. $D_{\rm v}\!<\!\!<\!D_{\rm h}$ and $\Omega\left(r,\theta\right)\approx{\overline\Omega}\left(r\right)$. Next, Fig.~\ref{Fig:Omega_prof} shows the internal rotation profiles and the differential rotation rate (defined as $-r\Omega'/2\Omega$, where $'$ denotes the radial derivative ${\rm d}/{\rm d}r$) for all the models at the same ages as previously (upper and lower panels respectively). Given the results on the total horizontal diffusivity $D_{\rm h}$ presented in \S~\ref{CompDh}, 
adding the vertical source for horizontal turbulence $D_{\rm h,v}$, has globally little impact on the rotation profiles, which remain strongly differential during the overall MS evolution. The ZAMS almost flat profile exhibited by model A2 (fast rotator, light green) reflects a strong coupling at the ZAMS for the fast rotators, that is nonetheless rapidly wiped out by wind braking later on the MS evolution. At the 
ZAMS, the only distinction that can be made between the different models is related to their initial spin rate. 
This is due to the shallow differential rotation rates, also at that time; the models keep the rotational properties they have been given initially, meaning that a fast (slow) rotator is still fast (slow) rotator. Moreover, the PMS contraction and the relatively high angular velocity in the early phases where magnetic braking is not yet at its maximum efficiency lead to a very high $D_{\rm h,h}$. Therefore, the impact of the new component \textbf{$D_{\rm h,v}$} is negligible at this phase and the four models initialised with the same rotation rate are superimposed. At the age of the Sun and later at the end of the MS, we cannot distinguish anymore between the different rotators in terms of surface rotation. 
However, the effects of the different new turbulence prescriptions are now noticeable on the rotation profiles below the convective envelope. When including $D_{\rm h,v}$ (if active for $R_{\rm e}>R_{\rm e;c}$ for $R>0.5R_\star$), a slightly more important extraction of angular momentum is observed
 below the envelope because of the more efficient transport from the inside of the star to the surface driven by the enhanced meridional circulation sustained by the additional horizontal turbulent diffusion (Appendix \ref{appendix:TL}). 
As seen on the bottom panel of Fig.~\ref{Fig:Omega_prof}, the external layers of the radiative core are the place where the radial differential rotation rate is the more important. This is due both to the radius increasing and to the torque applied at the surface by the magnetised stellar wind. The vertical shear instability is then more able to develop and contributes to the horizontal turbulent diffusion coefficient (see Fig.~\ref{Fig:Hor_diag}).
Also, one can note that the rotation profile is independent of the chosen characteristic timescale for turbulence ($\tau$). Indeed, models A, B and C lead to very similar rotation profiles at the age of the Sun and at the end of the MS.

The rotation profile flatness obtained with \citet{MPZ2004} (model D) at the ZAMS is associated to a very strong meridional circulation triggered by the contraction during the PMS. This is especially true in the case of the fast rotating model since in their expression, $D_{\rm h,h}$ depends at the first order on the rotation rate.

The differential rotation obtained with the new prescriptions can also be seen in Fig.~\ref{Fig:Diag_Bouvier} where we draw on the top panel the evolution of the normalised core to the surface differential rotation (i.e. the difference between the mean rotation rate of the radiative core, $\Omega_{\rm rad}$, and the surface rotation, $\Omega_{\rm s}$, as defined in \citet{Amardetal2016} normalised by their sum) as a function of time. A value close to one indicates a very strong gradient of rotation between the radiative core and the convective envelope of the star, while $\Delta\Omega = 0$ means that the model rotates as a solid body. This diagram has to be read in parallel to the bottom panel which shows the evolution with time of both the surface angular velocity (gyrotracks) and the mean radiative core angular velocity. Predictions of the models are compared with rotation periods observed in stars with masses between 0.9 and 1.1~M$_{\odot}$ in open clusters over a large age range collected by \citet[][and references therein]{Bouvier2014}.
Note that the three models associated to different initial rotation rates reproduce fairly well the statistical data points for surface rotation with time, in a similar way as in \cite{Amardetal2016}. Even though the vertical shear turbulence prescription is different, it has little effects on the global transport of angular momentum, hence on the surface rotation.

The same prescription for the extraction of angular momentum by the wind is used for all the models (see \S~\ref{stellarmodels}). Therefore, the surface and the core rotation rates evolve differently only depending on the way we describe the horizontal turbulence. We see that the global evolutions of both the surface and the core rotation rates are almost unchanged from one model to the other. Indeed, even though the transport generated by the added term to the horizontal turbulent diffusion can be very high in amplitude locally, it is active only when the vertical shear instability is triggered. For a MS solar-mass star, this is the case for approximately 10\% of the lifetime, and this fraction decreases with the strength of the stellar-wind torque applied at the surface that generates radial differential rotation. 

While the horizontal turbulent transport induced by 3D motions triggered by the vertical shear instability can have a very high amplitude locally (for instance below the convective envelope), it is a self-regulated mechanism. Indeed, if the vertical shear is unstable and leads to important $D_{\rm h}$, an efficient meridional circulation is triggered that weakens the radial differential rotation (see Appendix \ref{appendix:TL}) because of the advection of angular momentum towards the surface to values below the threshold needed to sustain the vertical shear instability and the source of the associated strong horizontal turbulent transport thus vanishes. 
This self-regulation is the cause of the oscillations of the surface angular velocity that appear in the lower panel of Fig.~\ref{Fig:Diag_Bouvier} for ages larger that 600 Myrs approximately.
In addition, while the potentially strong horizontal turbulent transport and the induced efficient advection of angular momentum {\it locally} flattens the rotation profile, it also steepens the gradient of angular velocity on each side of the turbulent region because of the spatial discretisation that is inherent to any stellar evolution numerical code. This enhances the shear with the neighbouring layers that in turn triggers the turbulence during the next time step. That transient phenomenon is responsible for the radial oscillations that are observed on Figs~\ref{Fig:Hor_diag},\ref{Fig:Anis},\ref{Fig:Deff}, and \ref{Fig:Omega_prof} below the convective envelope on the main sequence when $D_{\rm h,v}$ is locally several orders of magnitude higher than $D_{\rm h,h}$. A smoother transition would likely be possible with a {\it non-local} instability criterion \citep[see e.g.][]{GagnierGaraud2018}.

As a partial conclusion, additional transport mechanisms such as internal gravity waves, magnetic fields or other instabilities (we refer the reader to the introduction for relevant references) are therefore still necessary to explain the efficient extraction of angular momentum from the core of low-mass stars during their PMS and MS. The case of subgiant and red giant stars, which have been successfully sounded thanks to space-based asteroseismology, should also be examined.

\subsection{Case of red giant stars illustrated with a 1.25 M$_\odot$ model}
\label{1.25model}

\begin{figure}[ht]
\includegraphics[width=0.45\textwidth]{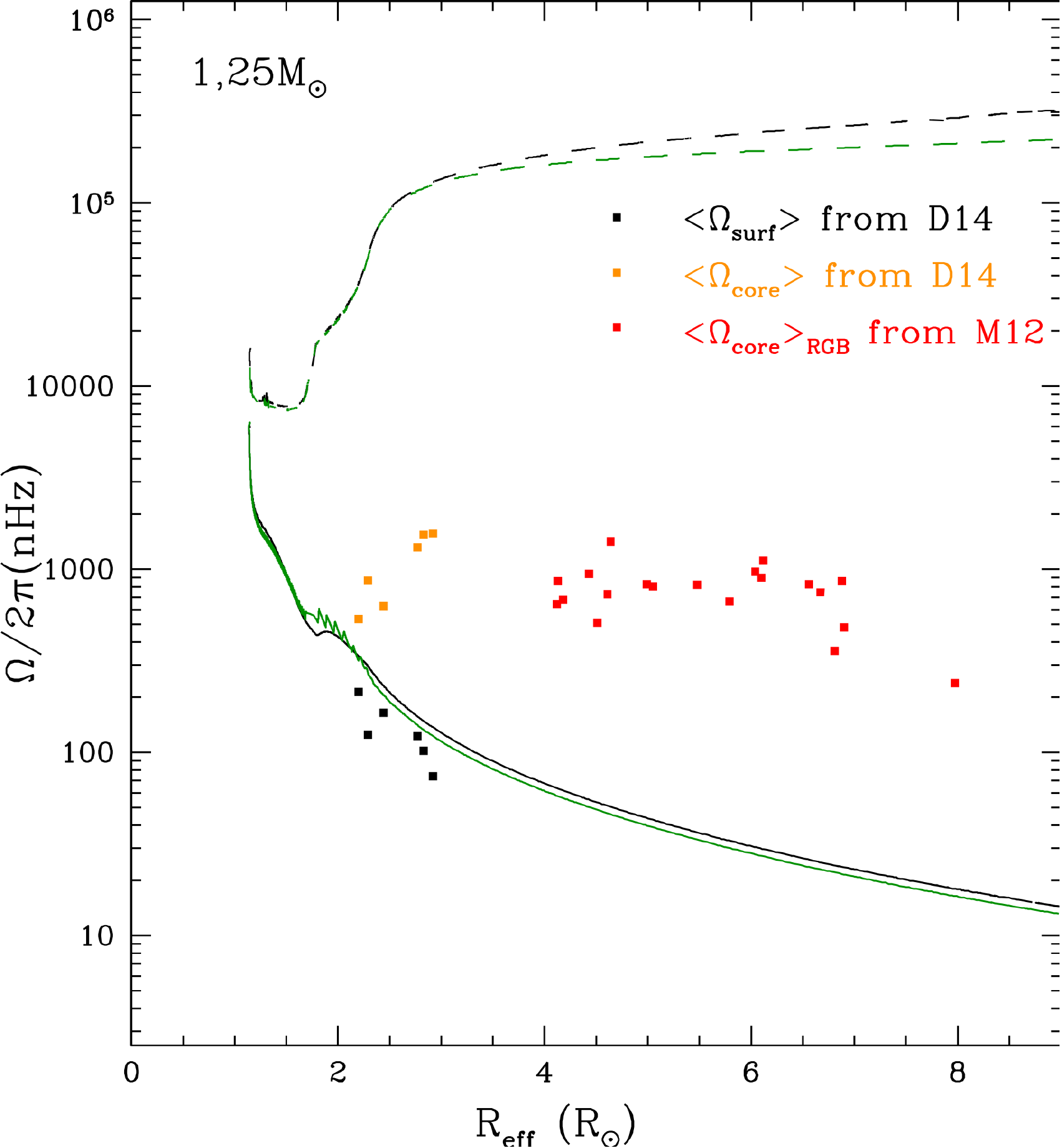}
\caption{Evolution of the surface and core rotation rates (solid and dashed lines respectively) as a function of the stellar radius that increases with time (MS, subgiant, and RGB correspond respectively to R$_{\rm eff}$/R$_\odot$ between $\sim$ 1.0 and 1.8, 1.8 and 3, and higher than 3). We show theoretical predictions for two 1.25M$_\odot$ models at solar metallicity computed with the prescription for $D_{\rm h}=D_{\rm h,h}$ by \citet{MPZ2004} (black) and the new prescription including both components of $D_{\rm h}=D_{\rm h,v}+D_{\rm h,h}$ for the case $\tau = 1/S$ (green). We compare the model predictions with relevant data for stars with asteroseismic masses between 1.1 and 1.4 M$_\odot$. For SGB stars we show surface and core rotation rates (black and orange dots) derived by \cite{Deheuvelsetal2014}. For RGB stars, we plot the core rotation rates derived by \cite{Mosseretal2012b} (red dots).}
\label{Fig:giant}
\end{figure}

The asteroseismic data obtained for the stellar core rotation from the mixed modes stochastically excited in sub-giants and red giants provide very good constraints for the transport of angular momentum in evolved stars \citep[e.g.][]{Becketal2012,Mosseretal2012b,Deheuvelsetal2012,Deheuvelsetal2014,Deheuvelsetal2015,Spadaetal2016,Gehanetal2018}. The observed stars have masses ranging between 1.1 and 1.4~M$_\odot$. We thus focus now on the intermediate mass of 1.25~M$_\odot$ for which we evolve models from the PMS to the RGB phase with the same assumptions and prescription for the loss of angular momentum through stellar winds as for the previous 1.0 M$_\odot$ models.
We saw in Sect.~\ref{subsec:RotEvol} that the time-scale associated with the turbulence ($\tau$) does not change drastically the results obtained for the rotation evolution. Therefore, to simplify the comparison, we only show here
two 1.25M$_\odot$ models with D$_{\rm h}$ computed following \citet{MPZ2004} only and the new one assuming that $\tau=1/S$ (same transport as model A)\footnote{Models with $\tau = 1/(2\Omega+S)$ and $\tau = 1/N_\Omega$ have been computed and are indistinguishable from the model computed with $\tau = 1/S$.}. 

Fig.~\ref{Fig:giant} shows the evolution of the surface rotation of these two models compared to the data given by \cite{Deheuvelsetal2014} for a few sub-giant stars. The braking is not calibrated to reproduce the surface velocity of such stars, therefore we do not expect to reproduce quantitatively the data. As such, the surface angular velocity of the models is too high. Two effects may be responsible for it : a lack of angular momentum extraction by the wind as well as a lack of transport in the radiative region of the star. While the first seems obvious, the latter needs a few explanations: a more efficient transport of angular momentum on the MS would lead to a smaller reservoir of angular momentum when the star begins to expand on the sub-giant branch, thus to a smaller angular velocity.

On Fig.~\ref{Fig:giant}, we also show the angular velocity of the rotating core of the same models. To be able to compare them with asteroseismic data, we compute the core rotation rate as the mass average of the angular velocity in the resonant cavity of gravity-dominated modes, that is in the region below the peak of the chemical Brunt-Väisälä frequency ($N^2_\mu = \displaystyle{\frac{g\varphi}{H_P}\nabla_\mu}$, with $H_P$ the pressure scale height, $\nabla_\mu$ the mean molecular weight gradient, and $\varphi=\left(\partial\ln\rho/\partial\ln\mu\right)_{P,T}$). As for the MS evolution, the rotational tracks are very close. The model with the new prescription shows a core rotating slightly slower as it is expected from a model with a more efficient coupling. However, this difference is far from sufficient to explain the asteroseismic data derived for the subgiants and red giants cores by \citet{Mosseretal2012b}. Hence again, other mechanisms such as propagative internal gravity waves \citep{TC08,Pinconetal2017}, mixed modes \citep{Belkacemetal2015a,Belkacemetal2015b},  or MHD instabilities \citep{Rudigeretal2015} should be invoked to potentially reproduce observations.

\section{Conclusion}
\label{sec:conclusion}
In this work, we derive new theoretical prescriptions for the anisotropy of the turbulent transport in differentially rotating stellar radiation zones. Extending the theoretical formalism derived by \cite{KB2012} to the case of rotating stably stratified flows with a vertical shear, we find that the ratio between the horizontal and the vertical turbulent transport scales as $N^4\tau^2/\left(2\Omega^2\right)$, where we recall that $N$ and $\Omega$ are the buoyancy and rotation frequencies, respectively, and $\tau$ is the time characterizing the source of the turbulence. This shows that if anisotropy increases with stratification, the Coriolis acceleration constitutes the restoring force in the horizontal direction, which was 
ignored in previously derived prescriptions. 

We propose here three physically-motivated expressions for $\tau$: $\tau=1/S$, $1/\left(2\Omega+S\right)$, and $1/N_{\Omega}$, where $S$ is the shear and $N_{\Omega}$ the epicyclic frequency. The first choice corresponds to the model for the turbulent transport induced along the radial direction by the vertical shear instability proposed by \cite{Zahn1992} and validated by recent direct numerical simulations \citep[][]{Pratetal2013,Garaud2016}. The second and third choices correspond to the introduction of the influence of rotation on the shear-induced turbulence. Their robustness should be validated in a near future by direct numerical simulations of rotating stably stratified flows with an unstable vertical shear in the range of parameters that corresponds to stellar regimes. These needed simulations, which are out of the scope of this work, will also allow us to improve the \cite{Zahn1992}'s turbulent model by taking into account the action of the Coriolis acceleration, of the chemical stratification and viscosity as in \cite{Pratetal2016}, and of the interactions with the horizontal turbulence \citep{TZ1997}.

Then, the turbulent transport coefficient in the horizontal direction, induced by the 3D turbulent motions sustained by the vertical shear-driven instability, scales as $Ri_{\rm c}\displaystyle\left(N/2\Omega\right)^2 K f(S,\Omega)$, where $\rm Ri_{c}$ and $K$ are the critical Richardson number and thermal diffusivity, respectively, and $f$ is a function that depends on the prescription chosen for $\tau$, when assuming the model proposed by \cite{Zahn1992} for the vertical turbulent transport coefficient. We identify that $f=1$ for $\tau=1/S$. 
We recall that all the previous prescriptions \citep{Zahn1992,Maeder2003,MPZ2004} have no explicit dependence on the entropy (and chemicals) stratification and do not take the action of the Coriolis acceleration into account, although these are the two restoring forces for turbulent flows in differentially rotating stellar radiation zones.

When applied to complete stellar evolution models of rotating low-mass, solar-metallicity stars, accounting for the feedback of the vertical shear in the horizontal direction ($D_{\rm h} = D_{\rm h,h}+D_{\rm h,v}$), the new prescriptions do not modify greatly the results previously obtained using the \citet{MPZ2004} prescription ($D_{\rm h} = D_{\rm h,h}$) for the internal and surface rotation rates. Although the horizontal turbulent transport induced by 3D motions triggered by the vertical shear instability can have a very high amplitude locally (for instance below the convective envelope of low-mass main-sequence stars), it is a self-regulated mechanism. Indeed, if the vertical shear is unstable and leads to important $D_{\rm h}$, an efficient meridional circulation is triggered that weakens the radial differential rotation to values below the threshold needed to sustain the vertical shear instability and the source of the associated strong horizontal turbulent transport thus vanishes. This also leads to a quenching of the vertical turbulent transport. As a consequence, this mechanism is not able to provide the extraction of angular momentum from stellar cores needed to explain their rotation rates in the Sun and in main-sequence, subgiant and red giant low-mass and intermediate-mass stars. A supplementary mechanism such as internal gravity waves or/and magnetic fields and their instabilities should thus be invoked. In this framework, this work demonstrates the great importance of pursuing the efforts to consistently model turbulent transport induced by rotation-driven hydrodynamical instabilities, that have a major impact on the structural and rotational evolution of stars. This strongly motivates future works to improve: (i) the physical description of the radial turbulent transport induced by the vertical shear instability with taking the action of the Coriolis acceleration, the effects of the chemical stratification and of viscosity \citep[][]{Pratetal2016}, and the interactions with the horizontal turbulence \citep{TZ1997} into account; (ii) the prescriptions for the horizontal and vertical turbulent transports induced by instabilities of the horizontal differential rotation. Their strengths have to be carefully evaluated in order to be able to discuss in a consistent framework the effects of other transport mechanisms of angular momentum such as internal gravity waves and magnetic fields.
\begin{acknowledgements}
We dedicate this work to the memory of our friend and mentor Jean-Paul Zahn, in the steps of whom we walk. Pursuing the path he opened is a great honor. 

We warmly thank the referees for their careful reading and constructive reports that allowed us to improve our work.

S.~M. and V.~P. acknowledge support by ERC SPIRE 647383 and FP7 SpaceInn grants. S.~M., V.~P. and A.~P. acknowledge support by CNES PLATO funding. C.~C. and L.~A. thank the Equal Opportunity Office of the University of Geneva. The authors acknowledge financial support from the French Programme National de Physique Stellaire PNPS of CNRS/INSU,  the grant ANR 2011 Blanc SIMI5-6 020 01 Toupies (Towards understanding the spin evolution of stars) and the Swiss National Science Foundation (SNF).
\end{acknowledgements}


\bibliographystyle{aa}  
\bibliography{Mathisetal16}



\begin{appendix}

\section{Horizontal averages}
\label{appendix:HA}

\subsection{The case of $\tau=1/\left(2\Omega+S\right)$}

The use of this prescription for the turbulent characteristic time requires to compute integrals of the form
\begin{equation}
I = \displaystyle{\int_0^\pi\frac{\sin^3\theta\,{\rm d}\theta}{\left(\displaystyle{\frac1x} + \sin\theta\right)^2}},
\end{equation}
which are also defined when $x>-1$.
The substitution $t=\tan\displaystyle{\frac{\theta}{2}}$ gives
\begin{equation}
I = 16x^2\int_0^\infty\frac{t^3{\rm d}t}{(1+t^2)^2\left[(t+x)^2+1-x^2\right]^2}.
\end{equation}
The partial fraction decomposition of the integrand is
\begin{equation}
\label{eq:dec}
\begin{aligned}
	-&\frac{1}{4x^3}\frac{1}{t^2+1} + \frac{1}{4x^2}\frac{t}{(t^2+1)^2} + \frac{1}{4x^3}\frac{1}{(t+x)^2+1-x^2}	\\
	&+ \frac{1}{4x^2}\frac{t}{\left[(t+x)^2+1-x^2\right]^2}.
\end{aligned}
\end{equation}

When $1-x^2>0$ (i.e., 
when $-1<x<1$), this leads to
\begin{equation}
I = 2\frac{
\begin{aligned}
&\left[(x-\pi)(1-x^2) + x\right]\sqrt{1-x^2}\\
&\ + (2-3x^2)\left[\frac{\pi}{2}-\arctan\left(\frac{x}{\sqrt{1-x^2}}\right)\right]
\end{aligned}
}{x(1-x^2)^{3/2}}.
\end{equation}

When $1-x^2<0$ (for $x>1$), the last two terms of Eq.~\eqref{eq:dec} can be further decomposed, and one finally obtains
\begin{equation}
I=\frac{2\left[(x-\pi)(x^2-1) - x\right]\sqrt{x^2-1} + (2-3x^2)\ln\left|\displaystyle{\frac{x-\sqrt{x^2-1}}{x+\sqrt{x^2-1}}}\right|}{x(x^2-1)^{3/2}}.
\end{equation}
These analytical results allow us to avoid to compute latitudinal averages in Eqs. (\ref{eq:Dh2}) and (\ref{eq:Dh3}) numerically and to speed up their evaluation.

\subsection{The case of $\tau=1/N_{\Omega}$}

The use of this prescription requires to compute integrals of the form
\begin{equation}
J = \int_0^\pi\frac{\sin^3\theta\,{\rm d}\theta}{\displaystyle{\frac1x} + \sin^2\theta},
\end{equation}
which are defined when $x>-1$.
The classical substitution $t=\cos\theta$ gives
\begin{equation}
J = \int_{-1}^1\frac{t^2-1}{t^2-\left(1+\displaystyle{\frac{1}{x}}\right)}{\rm d}t.
\end{equation}

When $1+1/x>0$, that corresponds to $x>0$, we use the partial fraction decomposition
\begin{equation}
\frac{t^2-1}{t^2-\alpha^2} = 1 + \frac{\alpha^2-1}{2\alpha}\left(\frac{1}{t-\alpha} - \frac{1}{t+\alpha}\right)
\end{equation}
with $\alpha^2 = 1+1/x$
to obtain
\begin{equation}
J = 2 + \frac{1}{x\sqrt{1+\displaystyle{\frac1x}}}\ln\left|\frac{1-\sqrt{1+\displaystyle{\frac1x}}}{1+\sqrt{1+\displaystyle{\frac1x}}}\right|.
\end{equation}

When $1+1/x<0$, that corresponds to $-1<x<0$, the partial fraction decomposition
\begin{equation}
\frac{t^2-1}{t^2+\alpha^2} = 1-\frac{1+\alpha^2}{t^2+\alpha^2}
\end{equation}
leads to
\begin{equation}
J = 2 + \frac{2}{x\sqrt{\left|1+\displaystyle{\frac1x}\right|}}\arctan\left(\frac{1}{\sqrt{\left|1+\displaystyle{\frac1x}\right|}}\right).
\end{equation}

\section{Validation of the shellular rotation assumption}
\label{appendix:shellular}

\begin{figure*}
\centering
\label{Fig:Rossby}
\includegraphics[width=0.8\textwidth]{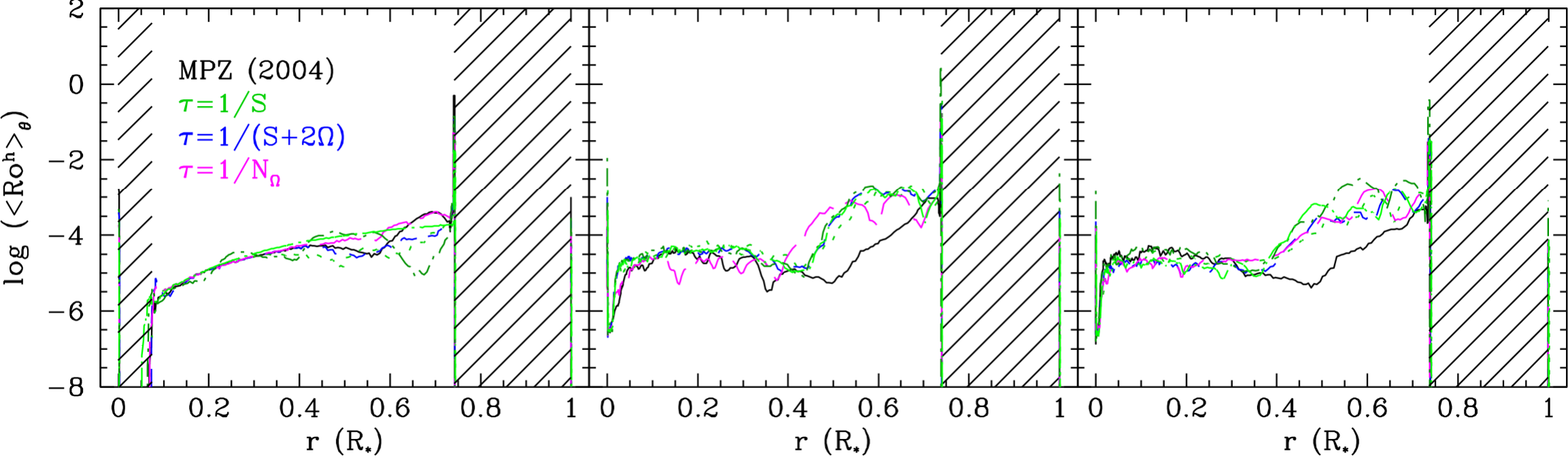}
\caption{Ratio between the horizontal variation of the angular velocity and the mean angular velocity on an isobar, expressed as a horizontal Rossby number, for the same models and color code as in Fig.~\ref{Fig:Hor_diag}.}
\end{figure*}

We recall the first-order expansion of the angular velocity introduced by \cite{MZ2004}: 
\begin{equation}
\Omega\left(r,\theta\right)={\overline\Omega}\left(r\right)+\Omega_{2}\left(r\right)Q_{2}\left(\theta\right)
\label{Eq:OmExp}\end{equation}
with $Q_{2}\left(\theta\right)= P_{2}\left(\cos\theta\right)+\displaystyle{\frac{1}{5}}$, where $P_2(\theta)$ is the second-order Legendre polynomial, and
\begin{equation}
{\overline \Omega}=\frac{\int_{0}^{\pi}\Omega\sin^3\theta\,{\rm d}\theta}{\int_{0}^{\pi}\sin^3\theta\,{\rm d}\theta}
\end{equation}
the so-called "shellular" rotation introduced by \cite{Zahn1992}, which is the average of the angular velocity on an isobar.

According to \cite{Zahn1992} and \cite{MZ2004}, the asymptotic solution for $\Omega_2$ can be expressed as :
 \begin{equation}
 \frac{\Omega_{2}}{\overline \Omega}=\frac{1}{D_{\rm h}}\frac{1}{5}r\left[2 V_{2}\left(r\right)- \alpha\left(r\right)U_{2}\left(r\right)\right]\quad\hbox{with}\quad\alpha=\frac{1}{2}\frac{{\rm d}\ln\left(r^2{\overline\Omega}\right)}{{\rm d}\ln r}
 \end{equation}
 when solving the equation for the transport of the angular momentum in the latitudinal direction assuming a short turbulent diffusion time-scale $r^2/D_{\rm h}$. Here, $U_{2}$ and $V_{2}$ are the $l=2$ radial functions of the vertical and latitudinal components of the rotation-driven meridional circulation repectively
 \begin{equation}
 {\vec {\mathcal U}}_{\rm M}\left(r,\theta\right)=U_{2}\left(r\right)P_{2}\left(\cos\theta\right){\vec e}_{r}+V_{2}\left(r\right)\frac{{\rm d}P_{2}\left(\cos\theta\right)}{{\rm d}\theta}{\vec e}_{\theta},
 \end{equation}
 which is expanded on Legendre polynomials. This sub-sonic large-scale flow verifies the anelastic approximation, i.e. ${\vec\nabla}\cdot\left({\rho}{\vec {\mathcal U}}_{\rm M}\right)=0$, where acoustic waves are filtered out, that leads to 
 \begin{equation}
 V_{2}=\frac{1}{6{\rho}r}\frac{{\rm d}}{{\rm d}r}\left({\rho} r^2 U_{2}\right).
 \end{equation}
 
With $D_{\rm h}$, $U_2$ and $\overline{\Omega}$ known, $V_2$ hence $\Omega_2$ can be computed at each time step and at each radius in the process of solving the angular momentum transport equation \citep{MZ2004}.\\

Then, we introduce the horizontal Rossby number (see also \cite{BC2001})
\begin{equation}
Ro^{\rm h}=\frac{u_{\perp}}{2\Omega l_{\perp}}\equiv \frac{1}{2\Omega}|\sin\theta\partial_{\theta}\Omega|.
\end{equation}
To follow this quantity using an 1-D stellar evolution code, we introduce its horizontal average over an hemisphere only
\begin{equation}
\left<Ro^{\rm h}\right>_{\theta}=\frac{\int_{0}^{\pi/2}Ro^{\rm h}\sin\theta\,{\rm d}\theta}{\int_{0}^{\pi/2}\sin\theta\,{\rm d}\theta}=\frac{3}{8}\frac{\Omega_2}{{\overline\Omega}},
\end{equation}
because of the anti-symmetry of $\sin^2\theta\,\partial_{\theta}Q_{2}\left(\theta\right)$ around the equator.
The horizontal Rossby number is thus given by the ratio between the horizontal differential rotation ($\Omega_2$) and the mean shellular rotation ($\overline\Omega$) similarly to the one introduced by \cite{SZ1992}.\\ 

To verify the shellular rotation assumption, we expect in the strongly stratified regime $D_{\rm v}\ll D_{\rm h}$ and thus $\Omega_{2}\ll{\overline\Omega}$ \citep[e.g.][]{Zahn1992}, i.e. $\left<Ro^{\rm h}\right>_{\theta}\ll1$. This is verified in our simulations where our new prescriptions are implemented as illustrated in Fig. \ref{Fig:Rossby}.
 











\section{Transport loop and strong horizontal turbulence}
\label{appendix:TL}

A larger $D_{\rm h}$ leads to an increase of the transport of angular momentum by the meridional circulation and of the horizontal diffusion of heat over an isobar (we refer the reader to the heat transport equation derived in \cite{MZ1998} (Eq. 4.36) and in \cite{MZ2004} (Eqs. 101-102)). As can be seen on  Fig.~\ref{Fig:Omega_prof}, which shows the rotation profile for the different \Dh{} prescriptions, the amplified transport leads to a locally weaker differential rotation (its evolution is discussed in details in \S \ref{subsec:RotEvol}) than with smaller horizontal turbulent diffusion coefficients. To understand the details of what is going on with a stronger horizontal transport, it is interesting to consider the hydrodynamical transport loop in stellar radiation zones as introduced by \cite{Rieutord2006} and \cite{Decressinetal2009}. First, the meridional circulation is mechanically driven by the wind and the shear-induced vertical transport (see Eq. 16 in \cite{Decressinetal2009}). Then, the circulation advects heat and because of the strong horizontal diffusion of entropy over the isobar, the temperature relaxes to a state where its fluctuation on an isobar is smaller than in the case with a weaker \Dh. This can be understood by looking at Eq. (19) in \cite{Decressinetal2009} and considering a simplified balance between heat advection and horizontal diffusion. Next, the radial gradient of rotation is deduced from the thermal wind balance given by Eq. (17) in \cite{Decressinetal2009}. A smaller temperature fluctuation leads to a smaller radial gradient of angular velocity. The vertical viscous shear-induced turbulent transport thus diminishes as well as the corresponding term driving the meridional circulation (Eq. 16 in \cite{Decressinetal2009}). This term (noted $U_{\rm v}$ in \cite{Decressinetal2009}) has an opposite sign compared to the term driven by the angular momentum extraction due to the stellar wind (noted $U_{\Gamma}$ in \cite{Decressinetal2009}). The amplitude of the meridional circulation thus increases in absolute value with a stronger advected flux of angular momentum towards the stellar surface and the transport loop is closed.

\end{appendix}

\end{document}